\documentclass[aps,prl,reprint]{revtex4-1}
\usepackage{graphicx}
\usepackage{dcolumn}
\usepackage{bm}
\usepackage[utf8]{inputenc}
\usepackage[T1]{fontenc}
\usepackage{mathptmx}
\usepackage{etoolbox}
\usepackage{lineno}
\usepackage{xcolor}
\usepackage{amsmath}
\usepackage{physics}

\makeatletter
\def\@email#1#2{%
 \endgroup
 \patchcmd{\titleblock@produce}
  {\frontmatter@RRAPformat}
  {\frontmatter@RRAPformat{\produce@RRAP{*#1\href{mailto:#2}{#2}}}\frontmatter@RRAPformat}
  {}{}
}%
\makeatother

\newcommand{\Sr}{\text{Sr}}
\newcommand{\He}{\text{He}}
\renewcommand{\Re}{\text{Re}} 

\begin{document}


\title{Momentum transfer in the outflow cycle of a Synthetic jet:\\ Comparison between a developed flow and an LE model}

\author{J.F. Hern\'andez-S\'anchez}
\author{F. Ordu\~na-Bustamante}
\author{R. Velasco-Segura}
\affiliation{Grupo de Ac\'ustica y Vibraciones, Instituto de Ciencias Aplicadas y Tecnolog\'ia, Universidad Nacional Aut\'onoma de M\'exico, Circuito Exterior S/N, Ciudad Universitaria, 04510, Mexico City.}

\date{\today}

\begin{abstract}
In the literature, flows produced by synthetic jets (SJ) have been studied extensively through experiments and numeric simulations. The essential physics of such a complex system has been simplified successfully to Lumped-element models in a wide range of conditions. LE models effectively predict the pressure in the cavity and the velocity in the neck of SJ. But, this does not comprise the complete dynamics of SJ. As soon as the flow starts separating from the neck of the SJ device, vortices and jets form at some distance downstream. These structures are the result of loosening the flow boundaries. Despite such a dramatic change, predictions of LE models remain unverified by measurements of the fully developed jet. We compared predictions of momentum transfer using an LE model with measurements of size and velocity of a fully developed jet/vortex detached from an SJ. Our SJ device operated with air as an active fluid. Comparing measurements and predictions, we found a constant difference for the higher sound pressures. However, the predictions and the measurements follow similar trends. Additionally, we found that the decay rate of the flow regime given by the relationship between the Reynolds and the Strouhal numbers differs significantly when the flow is studied within the neck and downstream the cavity.
\end{abstract}

\maketitle



\section{Introduction}


In a previous report, a synthetic jet device lifted quasi-steadily a sphere downstream its neck due to aerodynamic drag, resembling acoustic levitation\cite{Boullosa_AAuwA2010}. Such a drag develops at low frequencies and high sound pressure due to a periodic airflow around the sphere. According to reports, high-amplitude waves may induce air flows at the exit of pipes, a phenomenon known as acoustic streaming \cite{Fabre_ARFM2012}. The operating conditions in the previous report are consistent with regimes for vortex separation downstream the neck of the resonator and jet formation \cite{IngardLabate_tJotASoA1950}. Synthetic jets, a hybrid branch of physics combining acoustics and fluid dynamics, has been developed to study the characteristics of external flows generated by Helmholtz resonators. Synthetic-jet devices produce a periodic flow outwards the neck of the resonator, operating as intermittent pumps. Since the cavity of the resonator refills during the negative cycle of the oscillation, the external flow in synthetic jets does not need any addition of net mass \cite{Mohseni_CRCP2014}. Thus, synthetic jets are known in the literature alternatively as zero-net-mass-flux (ZNMF) jets. Despite not adding net mass, synthetic jets produce periodic bursts of momentum \cite{Krishnan_AIAAJ2009}. Synthetic jets dynamics has been investigated extensively through experiments, theory, and numeric simulations \cite{Glezer_ARFM2002}.

Experimental devices that produce synthetic jets are usually driven electromechanically by membranes and diaphragms vibrated with piezoelectric actuators~\cite{Glezer_ARFM2002}, or loudspeakers \cite{Gordon_PoF2004, Persoons_PoF2007}. Some experimental techniques have been implemented to visualize the flow, while others are applied to perform indirect measurements. Some of the visualization techniques were condensed by Alvi \& Cattafesta  \cite{Alvi_EPJSTT2010} (2010). For example, the smoke wire technique led to visualize streaklines  \cite{Brouckova_JV2015, Travnicek_AAIJ2012, Travnicek_TVSoJ2008}. Another technique, planar laser-induced fluorescence (PLIF), has been used to visualize the cross-section of a synthetic jet \cite{Alvi_EPJST2010, Ahmed_ETFS2009, Cater_JFM2002, Clark_IMEC&E2008, Gordon_PoF2004, Xia_JoMESc2012}. Thus, the mixing properties of synthetic jets have been estimated by measuring the concentration of dye \cite{Gordon_PoF2004}.The instantaneous velocity field is estimated after analyzing high-speed recordings with image analysis methods. One of these methods is Particle Image Velocimetry (PIV). In synthetic jets, PIV has been widely used to estimate the instantaneous velocity fields \cite{Amitay_PoF2006, Cater_JFM2002}. This technique has been used for other estimations, such as the mean velocity field \cite{Chaudhari_ET&FSc2009, Cicca_FDR2007, Lindstrom_AIAAJ2019, Shuster_PoF2007, Xia_AIAA2016} and its fluctuations \cite{Cater_JFM2002, Chaudhari_ET&FSc2009, Greco_JFM2017}, the mean vorticity \cite{Amitay_PoF2006, Lindstrom_AIAAJ2019, Xia_AIAA2016, Xu_ET&FSc2018}, the turbulence intensity \cite{Chaudhari_ET&FSc2009, Shuster_PoF2007}, the Reynolds stress \cite{Cicca_FDR2007, Wang_EJoMB2010} and velocity profiles.
Aside from the visualization techniques, hot-wire anemometry (HWA) has been also used to estimate the velocity of the flow \cite{Chaudhari_ET&FSc2009, Gil_ET&FSc2016, Kordik_ET&FSc2017, Lindstrom_AIAAJ2019, McGuinn_ET&FSc2013, Tesar_S&A2009, Travnicek_AAIJ2012, Travnicek_JV2015, Travnicek_TVSoJ2008, Vargas_AIAA2006, Xia_AIAA2016}.

Mainly two active fluids have been used in SJ devices: air \cite{Amitay_PoF2006, Brouckova_JV2015, Chaudhari_ET&FSc2009, Gil_ET&FSc2016, Greco_IJoH&FF2013, Guo_JAAIA2006, Jabbal_TAJ2006, Kordik_ET&FSc2017, Lindstrom_AIAAJ2019, McGuinn_ET&FSc2013, Tesar_S&A2009, Travnicek_AAIJ2012, Travnicek_JV2015, vanBuren_PoF2016, Vargas_AIAA2006, Xia_PoF2015}, and water \cite{Cater_JFM2002, Cicca_FDR2007, Clark_IMEC&E2008, Gordon_PoF2004, Greco_JFM2017, Shuster_PoF2007, Wang_EJoMB2010, Xu_ET&FSc2018}. Although, aqueous solutions of sugar and silicone oil have also been used as active fluids to further explore the effect of density and viscosity \cite{Xia_JoMESc2012}.

In synthetic jets, the stroke length is defined as the ratio between a velocity scale and the angular frequency
\begin{equation}
	L_s = \frac{u_\textrm{0}}{2\pi f}.
	\label{eq:StrokeLength}
 \end{equation} 
In some studies, $u_\textrm{0}$ is estimated as the momentum flow velocity \cite{Cater_JFM2002, Gordon_PoF2004, Greco_JFM2017}, while in others it scales as the stroke length times the frequency, $u_\textrm{0} \sim L_s f$ \cite{Amitay_PoF2006, McGuinn_ET&FSc2013, Shuster_PoF2007}. Here, the proportionality factor used is either $1$ or $2\pi$, depending on whether the angular frequency is used instead of the frequency. In a few cases, the velocity scale $u_\textrm{0}$ has been estimated from measurements at some distance downstream the neck of the resonator \cite{Chaudhari_ET&FSc2009, Gil_ET&FSc2016}. The stroke length plays a significant role in the formation of the external jet. In the investigation of Clark \textit{et al}. \cite{Clark_IMEC&E2008} (2008), the stroke length was not enough to reach the maximum circulation a vortex may transport and thus wakes did not develop. Jet formation takes place once the stroke length becomes significantly larger than the length of the neck of the resonator ($l_n$) and the outflow velocity becomes sufficiently high. Thus, comparing the stroke length with the geometry of the neck gives information about the dynamics of the jet. The dimensionless stroke length can be defined as the ratio between the stroke length  and the length of the neck $l_n$
\begin{equation}
	L_{s}^* = \frac{L_s}{l_n} =  \frac{u_\textrm{0}}{2 \pi f \; l_n}.
\end{equation}	
Commonly, the inverse of the dimensionless stroke length is defined as the Strouhal number, Sr = $1/L_{s}^*$. In this manuscript, the modified Strouhal number is defined in the neck instead as
\begin{eqnarray}
	\textrm{Sr}_n = \frac{f l_n}{u_\textrm{0}} \left(\frac{d_n}{l_n}\right) = \frac{f d_n}{u_\textrm{0}},
	\label{Eq:StrouhalNeckU0}
\end{eqnarray}
where $d_n$ is the inner diameter of the neck. The ratio $d_n/l_n$ is a dimensionless number summarizing the neck geometry. As defined in Equation~\eqref{Eq:StrouhalNeckU0}, the modified Strouhal number may be interpreted as a dimensionless frequency. Variations of the Strouhal number in synthetic jets have been previously investigated~\cite{Lee_AIAA2002, Tesar_S&AA2006,  Tesar_S&AA2010}. But in other investigations, the Strouhal number has remained fixed \cite{Greco_ET&FSc2016, Kral_AIAAJ1997}. Defined as in Equation~\eqref{Eq:StrouhalNeckU0}, the Strouhal keeps a close relationship with the formation number~\cite{Gil_JFS2018}. The formation number is defined as the stroke length divided by the hydraulic diameter, $L_s/d_H$. In some cases, the hydraulic diameter is smaller than the neck diameter, $d_H < d_n$. The effects of varying the formation number have been investigated previously \cite{Koso_JoS&T2014, Mohseni_OE2006, Mohseni_CRCP2014, Greco_ET&FSc2016}. The criterion for jet formation was discussed by Mohseni~\cite{Mohseni_OE2006}, where the maximum thrust is obtained at $L_s/d_H \approx$ 4. This result is consistent with previous findings in single stroke experiments \cite{Gharib_JFM1998, Rosenfeld_JFM1998, Palacios-Morales_AM2013}. However, at a high Stokes number, the formation criterion is more intricate \cite{Travnicek_AAIJ2012}. Furthermore, for different regimes, the criterion for formation depends on both the Strouhal and the Reynolds numbers simultaneously. A detailed discussion on this subject can be found in Brou{\v{c}}kov{\'a} \textit{et al.} \cite{Brouckova_MS&SP2016} (2016).

Despite the importance of the Strouhal number, its role is not enough to fully explain the dynamics of synthetic jets~\cite{Brouckova_MS&SP2016}. The role of viscosity in the formation and detachment of vortices has been discussed extensively for acoustic streaming. Viscosity was introduced initially as part of the solution to the wave equation~\cite{Eckart_PR1948}. This theory was especially important for ultrasonic frequencies~\cite{IngardLabate_tJotASoA1950}. For audible frequencies and high amplitudes, Ing\r{a}rd \& Labate~\cite{IngardLabate_tJotASoA1950} (1950) also verified that viscous effects lead to the formation of vortices and jets downstream the neck. The Reynolds number compares inertia with viscosity effects. Thus, the Reynolds number typically describes synthetic jets, in addition to the Strouhal number. The Reynolds number in the neck of the resonator is defined commonly as
\begin{equation}
	\textrm{Re}_n =  \frac{u_\textrm{0} d_n}{\nu},
\end{equation}
where $\nu$ is the air kinematic viscosity, and $d_n$ is the diameter of the neck. Generally, the Reynolds number establishes the flow regime in the neck of synthetic jets. The effect of variations in the Reynolds number has been investigated on some studies \cite{Amitay_PoF2006, Brouckova_JV2015, Cater_JFM2002, Chaudhari_ET&FSc2009, Gil_ET&FSc2016, Gil_JFS2018, Gordon_PoF2004}, while its value has  been kept with little variation by others \cite{Gilarranz_JFE2005, Greco_JFM2017, Koso_JoS&T2014, Kral_AIAAJ1997, Krishnan_EJoM2010, Lee_AIAA2002, McGuinn_ET&FSc2013, Pavlova_JoHT2006, Shuster_PoF2007}. Guo \& Zhong \cite{Guo_JAAIA2006} (2006) focused on synthetic jets in the micro-scale. Such a study led to relatively small Reynolds numbers, between 5 and 140. In particular, Guo \& Zhong \cite{Guo_JAAIA2006} (2006) showed there is a linear relationship between the normalized circulation and Reynolds number. Here, the circulation is normalized by dividing by the viscosity. In acoustic waves passing through perforated plates, the effects of viscosity have been considered via the Shear number instead \cite{Temiz_JoSV2016}.

Since the dynamics of synthetic jets is determined mainly by the Reynolds and the Strouhal numbers, alternatively defined as the dimensionless stroke length~\cite{Jabbal_TAJ2006}, both parameters may characterize them. The use of these two numbers has led to maps of the parametric space \cite{Cater_JFM2002, Gordon_PoF2004, Shuster_PoF2007, Travnicek_AAIJ2012, Travnicek_JV2015, Xia_JoMESc2012}. These maps describe different operational regimes of synthetic jets. Therefore, the points in these maps are scattered and do not typically converge to master curves. Hybrid synthetic jets, devices without moving parts, are an exception to this rule. Here it has been reported that the Reynolds and a modified Strouhal number scale as Re $\sim$ 1/Sr ~\cite{Tesar_S&AA2006}. Such a scaling rose after multiplying the Strouhal number by the ratio of the loop length and the diameter of the supply nozzle. In this investigation, the Reynolds and the Strouhal numbers are independent parameters.

Greco \textit{et al}.~\cite{Greco_IJoH&FF2013} (2013) reported that the vortex velocity decreases once it has detached from the nozzle, decreasing more with a larger distance while its size increases. Vargas \textit{et al}.~\cite{Vargas_AIAA2006} (2006) confirmed this result. This finding is not limited to velocity. Amitay \& Cannelle \cite{Amitay_PoF2006} (2006) found that the momentum expelled per cycle decays with the axial distance from the nozzle, conforming to an exponent of about -1.1. This is consistent with the findings of Cater and Soria~\cite{Cater_JFM2002} (2002) who found that synthetic jets have spreading rates and velocity gradients larger than continuous jets. The transition from laminar to a turbulent synthetic jet was investigated by Abraham \& Thomas~\cite{Abraham_CF2009} (2009). Laminar and turbulent synthetic jets display different decay rates of the local maximum velocity~\cite{Krishnan_EJoM2010}. 

Jabbal \textit{et al.}~\cite{Jabbal_TAJ2006} (2006) measured fluxes of mass and momentum as a function of the Reynolds number. Here, they propose two quantities as references of momentum transfer rate: (1) the product of the density, the driving frequency square, the peak-to-peak displacement square, and the neck diameter square, $\rho_0 f^2 \Delta^2 d_n^2$, and (2) the density times the kinematic viscosity square, $\rho_0 \nu^2$. Using the first quantity under-estimated the experimental results, while the second resulted in over-prediction. In contrast, normalizing the vortex circulation with the viscosity led to a good comparison between the experiments and the prediction. Kord{\'\i}k \textit{et al.} \cite{Kordik_ET&FSc2017} (2017) also estimated momentum fluxes as a function of the ratio between the driver and the nozzle diameters.
The external flow of synthetic jets has been used in several applications such as active flow control \cite{Persoons_PoF2007}, design of heat sinks \cite{Mahalingam_JEP2005}, electronic cooling \cite{Pavlova_JoHT2006}, mixing control~\cite{Davis_ASM&E1999}, air~\cite{Parviz_MT2005}, and underwater propulsion \cite{Mohseni_OE2006}, aerosol manipulation \cite{Qiao_PT2015}, and flow separation control \cite{Gilarranz_JFE2005}. Reports of synthetic jet devices with cavities of different volumes are in the literature for such applications. Some of these volumes are in increasing order 4.3~cm$^3$~\,\cite{Krishnan_AIAAJ2009}, 52~cm$^3$~\,\cite{Forner_22ICSV2015}, 115~cm$^3$~\,\cite{Persoons_PoF2007}, 710~cm$^3$~\,\cite{Ingard_TJASA1967}, 5,379~cm$^3$~\,\cite{Qiao_PT2015}.

Despite current advances, there are still some gaps in the literature to be addressed. Since the cycles of compression and expansion vary in space and time, synthetic jets should be modeled using the complete compressible Navier-Stokes equations. However, Lumped Element (LE) models have successfully simplified to ordinary differential equations some of the complex features of synthetic jets~\cite{Chiatto_S2017, Rienstra_AIAAJ2018, Singh_JoSV2014}. LE models may predict the pressure within the cavity and the particle velocity of the neck of Helmholtz resonators. However, structures such as vortices and jets begin to form downstream the neck as soon as there is flow separation. The formation of such structures requires that confined flows transform to free flows. Such a transformation may indirectly yield losses still unlinked to predictions of LE models. Such losses may be addressed by a quantitative assessment of the external flow. In particular, momentum within the neck and if the external flow should be conserved unless there are losses. Furthermore, SJ devices usually have relatively small cavities~\cite{Chiatto_S2017}. To our knowledge, the device in this study has the largest cavity than in previous studies for similar applications. Additionally, momentum estimations within the neck and at a certain distance downstream are typically considered exchangeable without further consideration.

In this investigation, we compared qualitative measurements of velocity and size of synthetic jets through an extensive range of conditions with numerical solutions of a Lumped Element (LE) model.


\section{Experimental setup and measurements}
\subsection{Experimental device}

\begin{figure*}
\centering
\includegraphics[width=1.95\columnwidth]{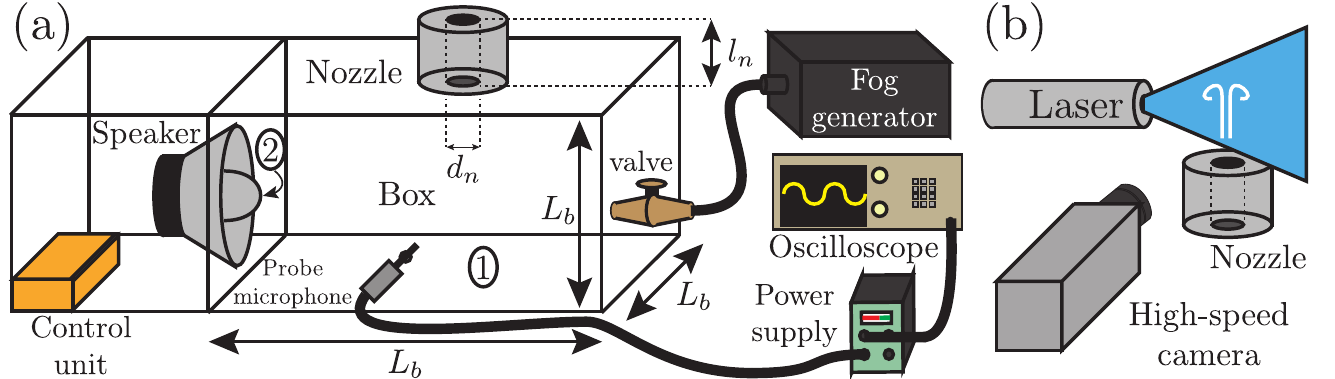}
\caption{\label{fig:Sketch}{(Color online)(a)~Sketch of the resonator device. A fog generator injects smoke into the right cavity to visualize the outer flow from the nozzle. (b)~A laser sheet illuminates the outflow from the nozzle while a high-speed camera records the experiments.}}
\end{figure*}

The experimental device has been previously described~\cite{Boullosa_AAuwA2010}. The device consists of an acrylic box with two inner cavities, as depicted in Figure~\ref{fig:Sketch}(a). A five-inch loudspeaker (Bumper 635IPS) is attached to the subdividing inner wall of the main cavity. The loudspeaker has a cone of a diameter of $d_s$ = 127~mm, a power rating of 250~W, and an electrical input impedance of 4~$\Omega$. The area of the speaker is estimated as $A_s = \pi d^2_s/4$. A homemade electronic control unit regulates the amplitude and frequency of the signal driving the loudspeaker. The control unit consists of a sine wave generator and an electronic amplifier that can drive the loudspeaker at frequencies up to 300~Hz and driving voltages up to 10.4~V. These parameters can be varied independently using two potentiometers. To obtain reproducible driving voltages, the amplitude potentiometer was adjusted to the same set of fixed positions during the experiments. The driving voltage and frequency are monitored in real-time using an oscilloscope (Tektronix MDO3032). Potentiometer settings correspond to the driving voltages shown in Table~\ref{Tab:VoltageValues}. Measurements within the cavity of the maximum RMS pressure and displacement are also reported here. The labels V$_i$ in this table are used throughout the manuscript to indicate the corresponding driving voltages in some of the figures and results. Note that the values of pressure and displacement vary with the driving frequency. Such variations were measured and can be seen as continuous lines in Figures~\ref{fig:DispacementVSFrequency} and \ref{fig:CavityPressure}. Table~\ref{Tab:VoltageValues} also shows the Sound Pressure Level (SPL) estimated at the maximum pressures.

\begin{table}
\centering
\caption{\label{Tab:VoltageValues}{RMS voltages supplied to the loudspeaker in the experiments at different driving regimes, maximum RMS loudspeaker cone displacements measured as a function of frequency, and maximum RMS values of internal sound pressure.}}
\mbox{}\\
\begin{tabular}{|c|c|c|c|c|} \hline
Label & V         & $\Delta x^\textrm{max}_s$ & $p^\textrm{max}_\textrm{rms, exp}$	& SPL$^\textrm{max}$ \\ 
	 & [volts]  & [mm]	                             &[Pa]							&[dB]\\ \hline
V$_0$ & $0.10  \pm 0.03$ & -- 			& 10, 20Hz&114.0 \\
V$_1$ & $2.15 \pm 0.03$ & 0.62, 20Hz & 131, 20Hz & 136.3\\
V$_2$ & $4.26 \pm 0.06$ & 1.24, 30Hz & 245, 20Hz & 141.8\\
V$_3$ & $6.43 \pm 0.10$ & 1.82, 30Hz & 350, 30Hz & 144.9\\
V$_4$ & $8.54 \pm 0.13$ & 2.39, 40Hz & 449, 40Hz & 147.0\\
V$_5$ & $10.43 \pm 0.42$ &2.94, 50Hz & 555, 50Hz& 148.9\\ \hline
\end{tabular}
\end{table}

The resonator opens to the environment through a nozzle tightly fitted on top of its upper face. The nozzle is made of PVC (polymerizing vinyl chloride). The right cavity in Figure~\ref{fig:Sketch}(a) has a cubic volume whose inner length is $L_b = 19$~cm. After considering the additional recess of the loudspeaker cone, the volume of such cavity is $V_c = 7.91\times10^{-3}~\mathrm{m^3}$. The inner diameter and the length of the nozzle are $d_n = 8$~mm and $l_n = 40.1$~mm. The external diameter of the nozzle is 40 mm, resulting in walls with a thickness of 16~mm. The inner edges of the neck have a chamfer radius of about 1~mm.

The Helmholtz resonant frequency of the right cavity is approximately given by the usual expression
\begin{equation}
  f_{\He} = \frac{c_0}{2\pi}\sqrt{\frac{A_n}{V_c l_n}},
  \label{Eq:HelmholtzResonance}
\end{equation}
where $c_0$ is the speed of sound, $A_n$ is the inner cross-sectional area of the nozzle, $A_n = \pi d_n /4$, and $l_n$ is the length of the nozzle. Assuming that the speed of sound is $c_0 = 343$~m/s results in a resonant frequency of $f_{\He} \approx 22$~Hz. If standard end corrections, flanged inside and unflanged outside, are added to $l_n$ to estimate an effective length, the resonance frequency estimate changes only by about 1~Hz. Such changes are small compared with the 5~Hz frequency steps used in our experiments.

\subsection{Loudspeaker displacement and velocity measurements}

We measured experimentally the oscillating displacement and velocity of the loudspeaker cone for the input range of operating frequencies and voltages. This measurement was performed using an accelerometer Endevco 27A11 and a signal conditioner Endevco 4416B. The accelerometer weighs 1 g, a mass about 100 times smaller than the speaker's cone. The accelerometer was attached to the center of the loudspeaker cone at the position labeled~(2) in Figure~\ref{fig:Sketch}(a). The signal from the accelerometer was acquired digitally by the oscilloscope, where RMS values were estimated. The acceleration was calibrated using a vibration calibrator Br\"{u}el \& Kj{\ae}r type 4294. This calibrator produces a reference RMS acceleration amplitude of $a_0 = 10~\mathrm{m/s}^2$ and a reference RMS displacement amplitude $\Delta x_0 = 10\ \mu\mathrm{m}$ at a frequency of 1000~rad/s ($\approx$159~Hz). This calibration resulted in a sensitivity of $10~\mathrm{m/s}^2$ every 102~mV.

\begin{figure}
\centering
\includegraphics[width=0.95\columnwidth]{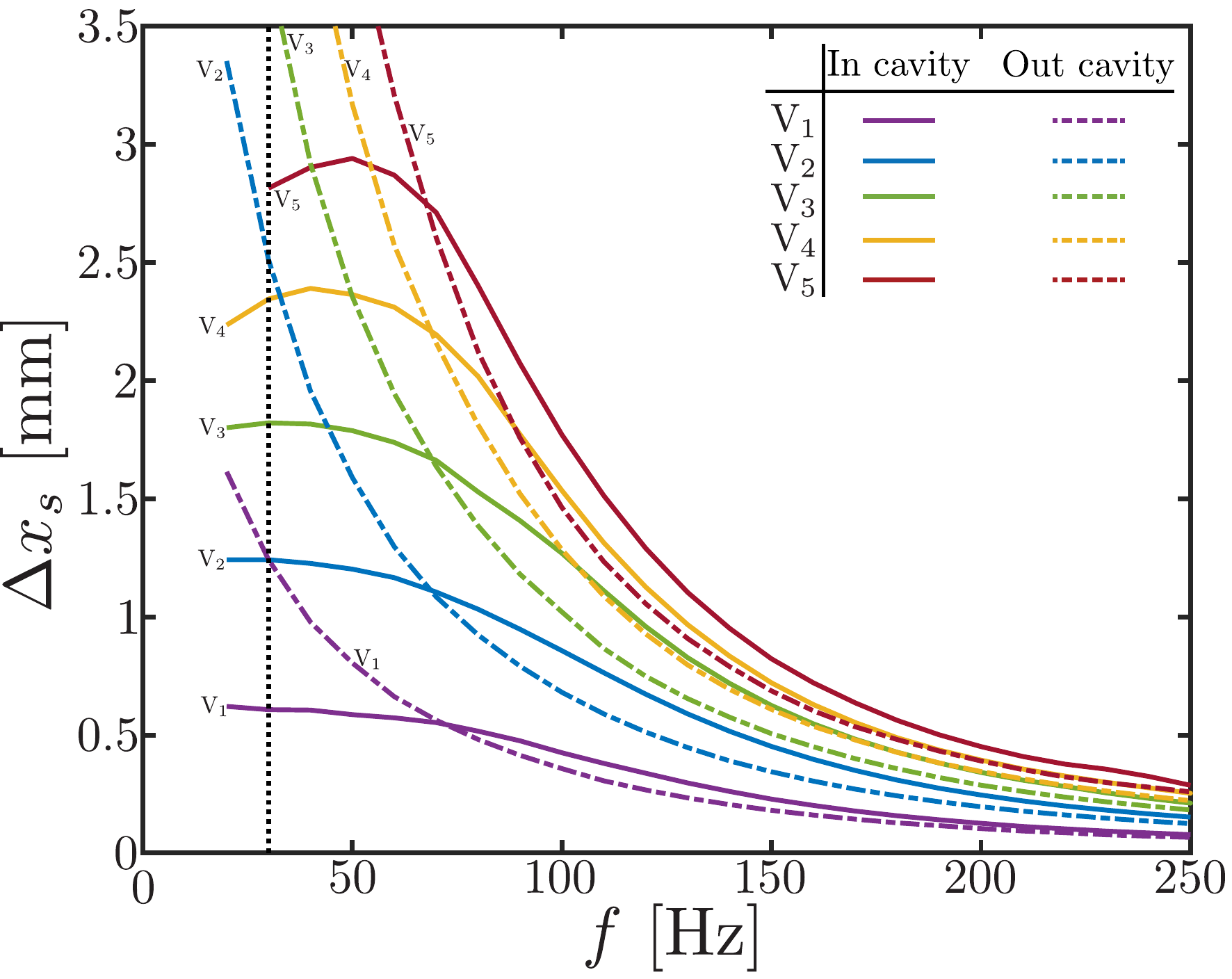}
\caption{\label{fig:DispacementVSFrequency}{(Color online) Measured RMS displacement of the loudspeaker cone at different driving frequencies. The colors and labels correspond to the different voltage labels in Table~\ref{Tab:VoltageValues}. The continuous lines correspond to measurements within the resonator and the broken lines correspond to measurements in a free environment.}}
\end{figure}

The mechanical response of the loudspeaker was studied with measurements of the RMS acceleration as a function of both the frequency and voltage. Such measurements were used to estimate the RMS velocity $v_s$ and displacement of the cone $\Delta x_s$ assuming the signal is sinusoidal, as
\begin{eqnarray}
  \label{Eq:RMSVelocity}
  v_s &=& \frac{a_s}{2\pi f},\\
  \label{Eq:RMSDisplacement}
  \Delta x_s &=& \frac{a_s}{4\pi^2 f^2},  
\end{eqnarray}
where $a_s$ is the RMS acceleration measured over the cone, and $f$ is the frequency of the driving signal. Equations~\eqref{Eq:RMSVelocity} and \eqref{Eq:RMSDisplacement} neglect the effects of higher harmonics. Figure~\ref{fig:DispacementVSFrequency} shows measurements of displacement $ \Delta x_s$ as a function of the driving frequency for the different input voltages. The continuous lines in Figure~\ref{fig:DispacementVSFrequency} correspond to the response of the loudspeaker within the cavity, while the broken lines correspond to measurements in a free unconfined environment. Thus, the maximum displacement of the loudspeaker within the cavity appears due to some resistance in confinement. For frequencies larger than 150~Hz, the displacement of the confined loudspeaker approaches the free response. The response of the loudspeaker was tested electrically, mechanically, and acoustically before the experiments.

Each curve in Figure~\ref{fig:DispacementVSFrequency} corresponds to a voltage level reported in Table~\ref{Tab:VoltageValues}. Thus, each voltage is associated with a curve of the driving amplitude of the loudspeaker. The lowest and highest voltages correspond to the $V_1$ and $V_5$ labels, respectively. The black dotted vertical line corresponds to the resolution limit found at 30~Hz. Acceleration values measured below this frequency limit have large uncertainty. Such a large uncertainty may be the result of the acceleration curves deviating significantly from a sine wave, thus conflicting with the assumption in Equations~\eqref{Eq:RMSVelocity} and \eqref{Eq:RMSDisplacement}. As the voltage is increased, the position of the maximum shifts towards higher frequencies. This frequency shift occurs because of the nonlinear response of the resonator and the loudspeaker. The values of displacement of the continuous lines in Figure~\ref{fig:DispacementVSFrequency} are the input conditions of the LE model described in Section~\ref{sec:LEM}.

\subsection{Sound pressure measurements}

\begin{figure}
\centering
\includegraphics[width=0.95\columnwidth]{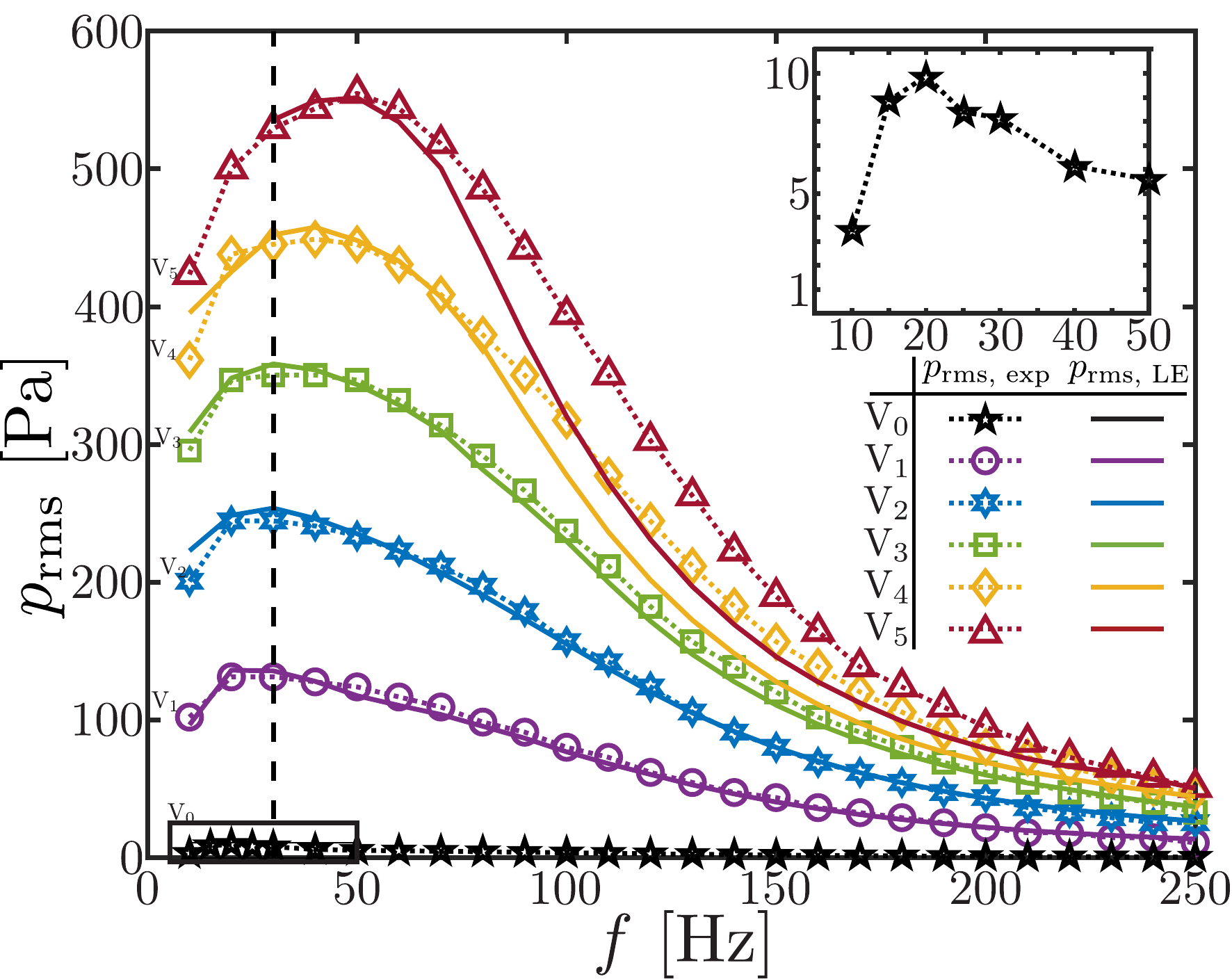}
\caption{\label{fig:CavityPressure}{(Color online) Sound pressure within the cavity of the resonator varying as a function of the driving frequency for different voltages. The continuous lines are the experimental measurements, broken lines are results from the simulation of the LE model.}}
\end{figure}

The RMS sound pressure $p_\textrm{rms, exp}$ was measured within the inner cavity of the resonator as a function of the frequency and voltage. The inner cavity is labeled as point~(1) in Figure~\ref{fig:Sketch}(a). Such measurement was performed using a probe microphone Br\"{u}el \& Kj{\ae}r type 4182, connected to a power supply Br\"{u}el \& Kj{\ae}r type 2807. The signal was acquired using the oscilloscope previously stated. The signal from the microphone was calibrated before each experimental session using a sound level calibrator Br\"{u}el \& Kj{\ae}r type 4230. This calibrator produces a signal at 1000~Hz and an RMS sound pressure of 1~Pa. 

The sound pressure $p_\textrm{rms, exp}$ was measured for the voltage levels in Table ~\ref{Tab:VoltageValues} and the range of driving frequencies. These measurements are seen as the open symbols on the dotted lines in Figure~\ref{fig:CavityPressure}. The maximum values of pressure $p^\textrm{max}_\textrm{rms, exp}$ are first reported in Table ~\ref{Tab:VoltageValues}. But, further detail seen in Figure~\ref{fig:CavityPressure} shows that the maximum pressure shifts toward higher frequencies with higher voltages. The solid lines in Figure~\ref{fig:CavityPressure} result from solving numerically the LE model (described in Section~\ref{sec:LEM}) using the displacement $\Delta x_s$, measured in Figure~\ref{fig:DispacementVSFrequency}, as an input condition. The output of this model is an estimation of the particle velocity on the neck of the resonator. These estimations of velocity are used to predict the momentum outflow from the neck of the resonator. To compare this prediction with experimental data, we estimated the momentum and RMS particle velocity from the properties of the external flow. The external flow consists of a series of vortices with different velocity and size. These two characteristics were estimated from the visualization. Hereafter, we describe the technique to visualize the flow.

The stars aligned with the dotted black curve in Figure~\ref{fig:CavityPressure} represent the measurements for the lowest voltage, $V_0$. The inset of this figure is a close-up to this curve, showing a maximum at about 20~Hz. This maximum corresponds to the Helmholtz resonance estimated from Equation~\eqref{Eq:HelmholtzResonance}. For higher voltages, the resonance shifts towards higher frequencies, mimicking the trend of the displacement in Figure~\ref{fig:DispacementVSFrequency}.

The displacement of the cone in Figure~\ref{fig:DispacementVSFrequency} and the acoustic pressure in Figure~\ref{fig:CavityPressure} shows a similar resonance shift toward higher frequencies. The experimental results and prediction of the LE model match reasonably well with one another. Since the displacement of the cone is not constant with frequency, the solid lines in Figure~\ref{fig:CavityPressure} are not strictly resonance curves. Similar to Figure~\ref{fig:DispacementVSFrequency}, the shift in frequency in Figure~\ref{fig:CavityPressure} occurs due to the nonlinear behavior of the system driven at high amplitudes.

\subsection{Visualization of the outflow}

\begin{figure*}
\includegraphics[width=1.95\columnwidth]{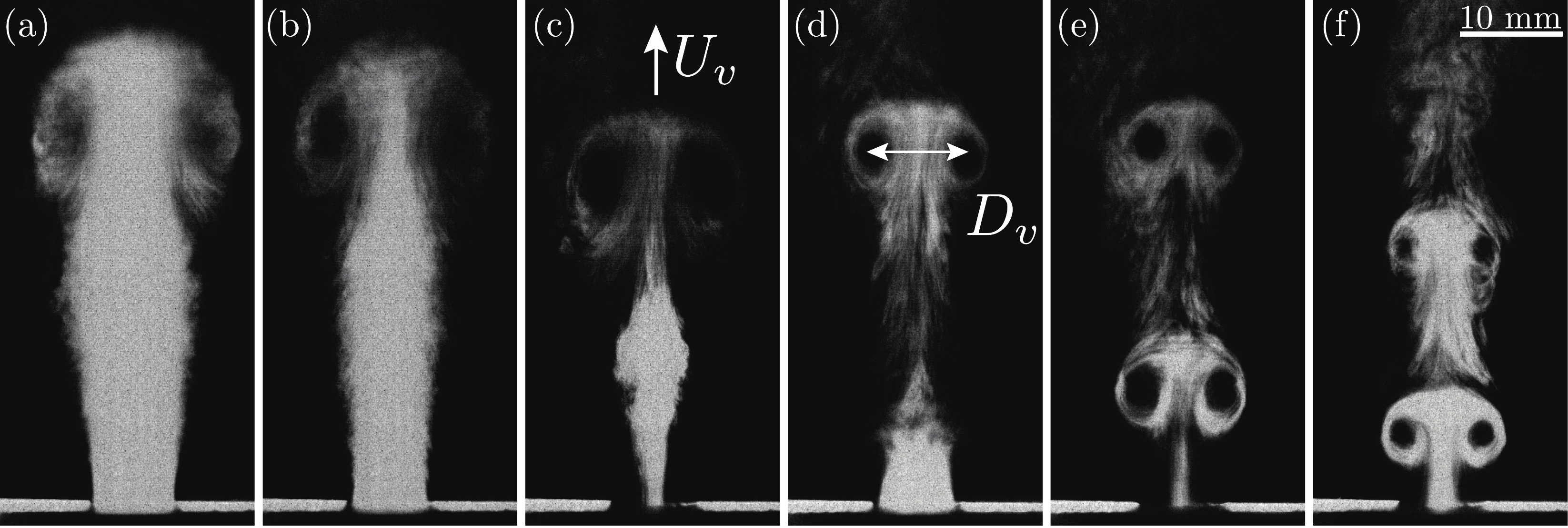}
\caption{\label{fig:Visualization}{Vortex rings visualized at voltage $V_4$. Visualization experiments at driving frequencies of (a)~40~Hz, (b)~80~Hz, (c)~110~Hz, (d)~130~Hz, (e)~140~Hz, (f)~150~Hz.}}
\end{figure*}

Smoke is injected into the cavity of the resonator with a fog generator (Dantec Dynamics 10D90P). The cavity is connected with the fog generator via a needle valve passing through one of the walls. The typical size of the particle tracers is between 1 and 3~$\mu$m. A narrow cross-section upstream the neck of the resonator was illuminated using a thin laser sheet, as shown in Figure~\ref{fig:Sketch}(b). The sheet was produced using a CW laser (S3 Arctic series from Wicked Lasers) attached to a cylindrical lens. This laser produced blue light at a single wavelength of 445~nm with a power of 3~W. These elements assemble the planar laser-induced fluorescence (PLIF) technique. We recorded the experiments using one of two high-speed cameras perpendicular to the flow: Motion Pro HS4 from Redlake, or Phantom V9.1 from Vision Research. The typical frame rate was 3000 frames per second (fps). Two Nikon lenses Micro Nikkor 60 mm with manual focus were mounted one on each camera. The panels in Figure~\ref{fig:Visualization}(a-f) show the evolution of vortices visualized for an increasing driving frequency and a fixed voltage $V_4$. The driving frequencies are in the caption of the figure. Each panel of Figure~\ref{fig:Visualization} results from the average image of about 10 vortices in the same phase of the cycle. At the low frequencies in Figure~\ref{fig:Visualization}(a-b), the vortex ring is followed by a long wake. As the frequency increases, the concentration of fog is reduced, as seen in Figure~\ref{fig:Visualization}(c). At even greater frequencies, the size of the vortex is reduced significantly. We observed that vortices are formed at some distance from the neck. If the negative cycle starts before the vortex is discharged, it is shredded and no external flow takes place. Thus vortices stop forming at frequencies above 150~Hz. In Figure~\ref{fig:Visualization}(e \& f), there is a single vortex discharged per cycle. In these two pictures, the velocity of the vortices is low enough to allow the formation of another vortex while the previous one is still in the field of view.

\subsection{Frequency of vortex formation and detachment}

\begin{figure}
\centering
\includegraphics[width=0.95\columnwidth]{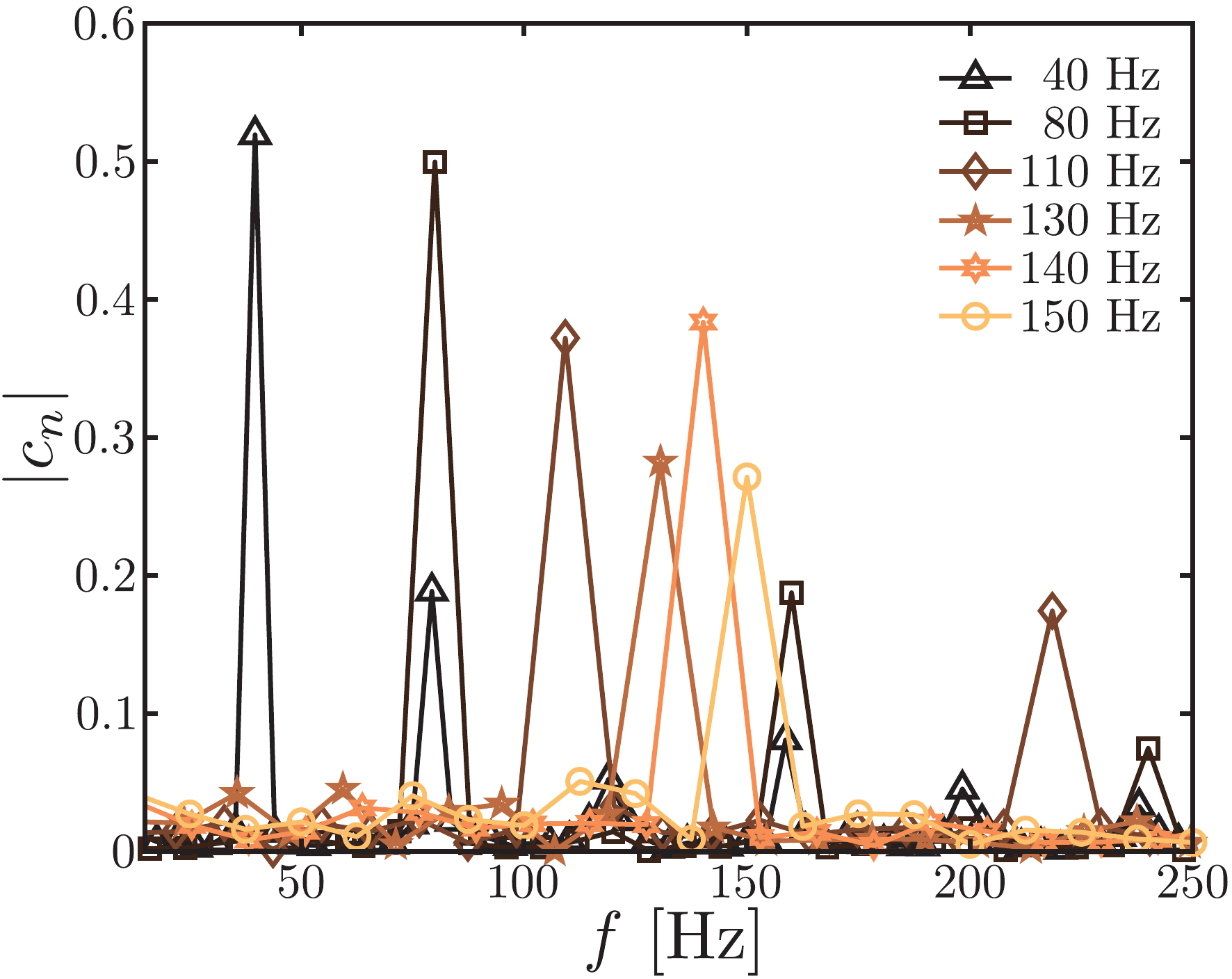}
\caption{\label{fig:fft}{(Color online) Coefficients of the fast Fourier transform of the fog concentration close to the neck seen as a function of the driving frequency. The symbols and colors correspond to the driving frequencies of 40, 80, 110, 130, 140, and 150~Hz.}}
\end{figure}

In the data reported by this manuscript, one vortex is discharged every cycle of the driver for the operational range of frequencies in Figure~\ref{fig:fft}. The frequency of vortex shedding was estimated as follows. A control area was defined above the neck. The mean change in the pixel intensity in the control area was estimated for successive frames, being proportional to the fog concentration. This concentration was normalized to the maximum value of the recording. At least ten cycles were recorded in each experiment. The Fourier coefficients $c_n$ were estimated using the fast Fourier transform on the temporal estimations of concentration. As an example, these coefficients are plotted as a function of frequency in Figure~\ref{fig:fft} for voltage $V_4$.

\subsection{Vortex velocity measurements}

\begin{figure}
\centering
\includegraphics[width=0.95\columnwidth]{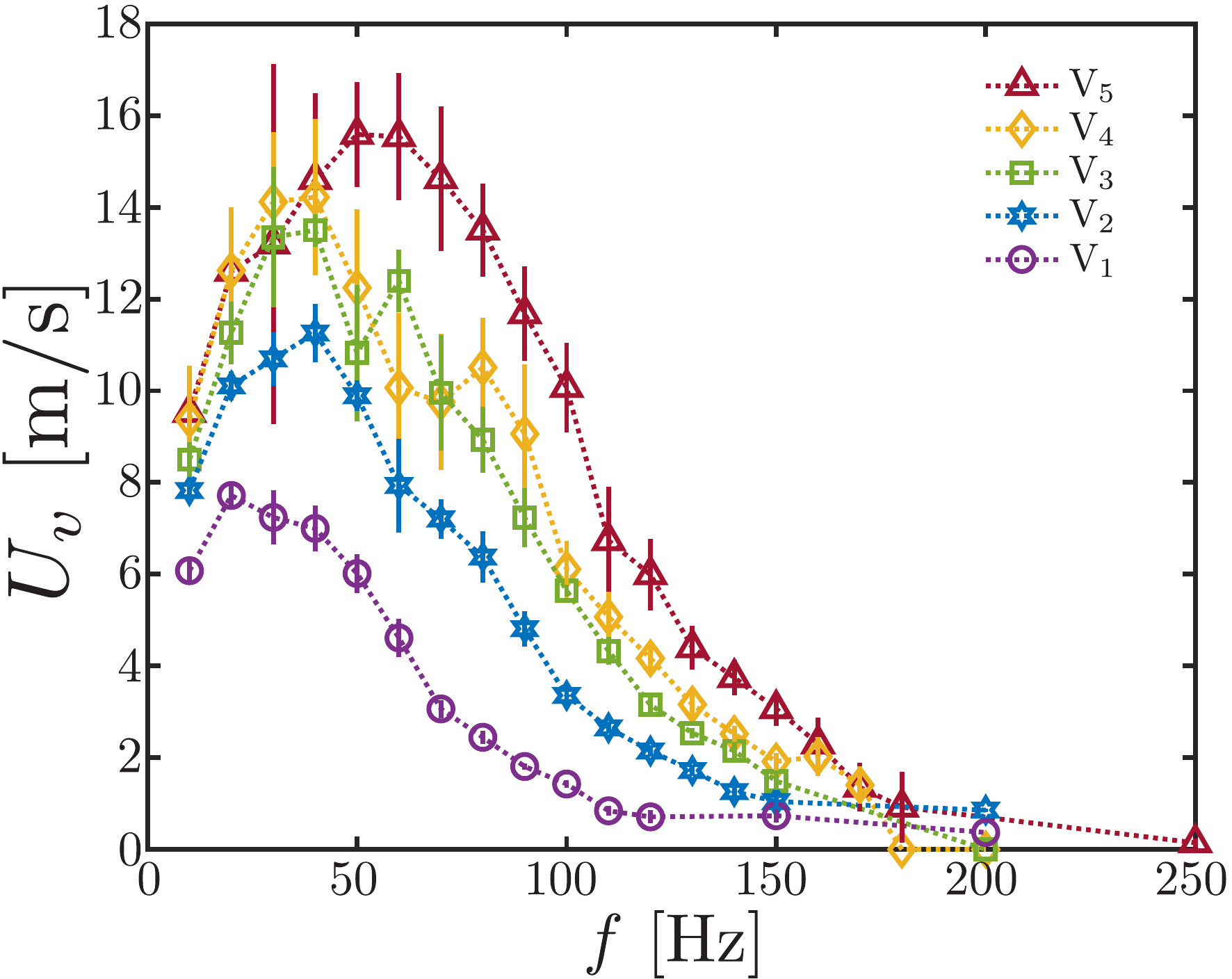}
\caption{\label{fig:VortexVel}{(Color online) Vortex vertical velocities as a function of the driving frequency for different voltages.}}
\end{figure}

The position of each vortex was measured by tracking its front in a sequence of frames of the recordings. Using linear regression, we fitted the positions as a function of time and estimated the average velocity of the vortex $U_v$ as the slope of the fit. The average velocity was estimated at about 4~cm downstream of the neck. Such an average estimation relies on the mean position of the vortex through a few frames, being more reliable than instantaneous velocity fields. Additionally, two high-speed recordings with two different conditions are shown in the Supplementary material. The open symbols aligned with the dotted lines in Figure~\ref{fig:VortexVel} show experimental measurements of the vertical velocity of the vortex $U_v$ as a function of frequency for different driving voltages. Vertical lines in Figure~\ref{fig:VortexVel} represent error bars estimated as the standard deviation of the velocity of at least ten vortices. Vortices usually form at a minimum distance upstream of the nozzle, at least 15~mm. The gap in the values between 200 and 250~Hz is a range of frequencies where vortices did not form. 

The velocity of the vortices increases with the driving voltage, reaching a maximum in mid frequencies. The corresponding frequency of such a maximum velocity shifts toward higher frequencies for higher voltages. This behavior is similar to the one of the sound pressure in Figure~\ref{fig:CavityPressure}. The vortex velocity and the sound pressure are both the result of the oscillation dynamics of the resonator and the loudspeaker. Such resonant dynamics are also displayed in the RMS displacement shown in Figure~\ref{fig:DispacementVSFrequency}.

RMS particle velocities obtained from the numeric simulations of the LE model, shown later in Figure~\ref{fig:ParticleVelocity}, are not immediately comparable with the vortex velocities. Therefore, velocities in Figure~\ref{fig:VortexVel} are not the particle velocities on the neck.

\subsection{Vortex size measurements}

\begin{figure}
\centering
\includegraphics[width=0.95\columnwidth]{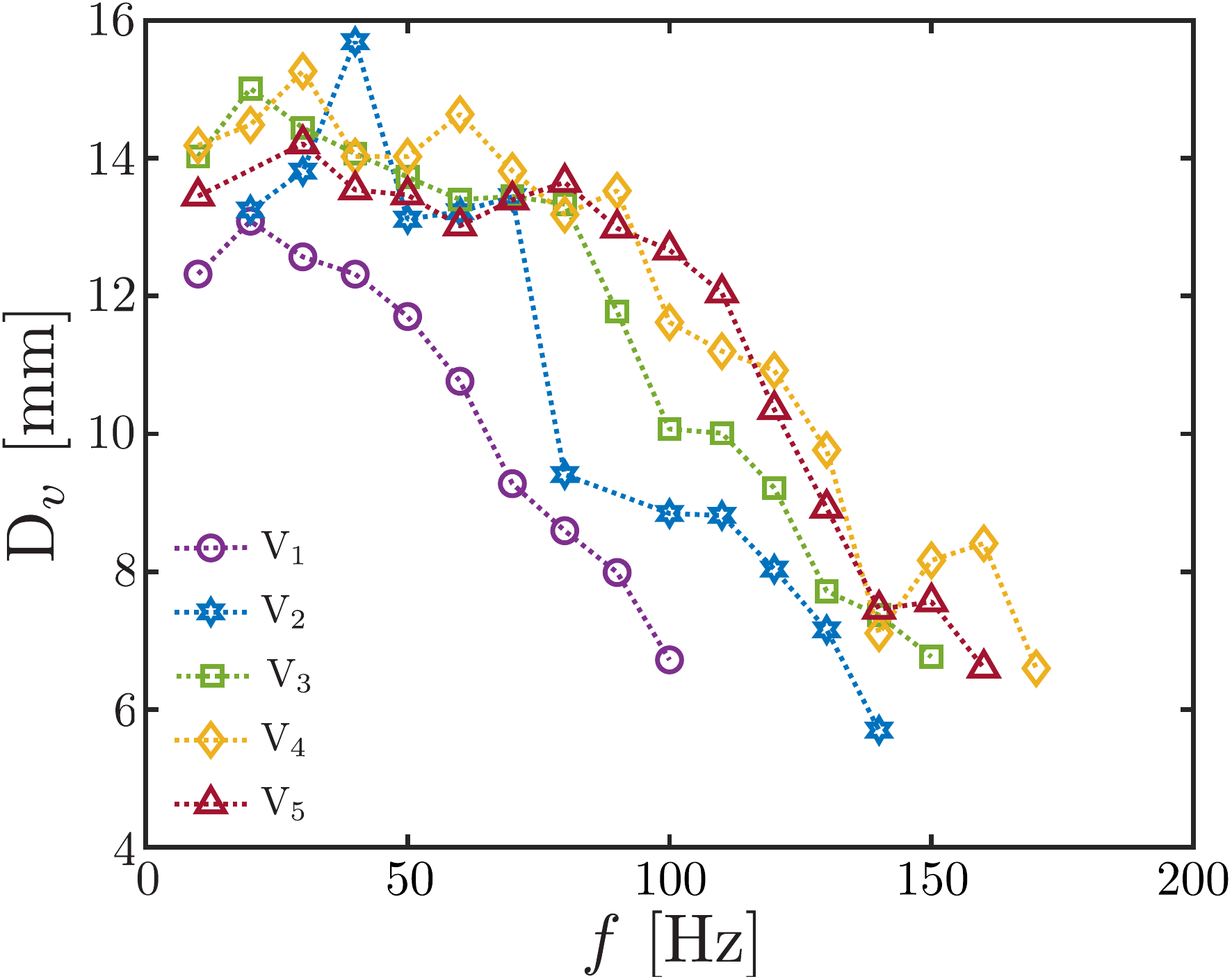}
\caption{\label{fig:VortexSize}{(Color online) Measurements of the diameter of the vortex as a function of the driving frequency for different voltages.}}
\end{figure}

The diameter of the vortices was estimated as the distance between the centers of the ring, seen as a two-point arrow in Figure~\ref{fig:VortexSize}(d). The mean diameter of the vortex, $D_v$, was estimated from the average image of about ten cycles at the same point in the cycle. Since the center of the vortex is not clearly seen in the average image, we have developed the procedure in the Appendix to estimate as precisely as possible. Essentially, the procedure consists on closing up to the two blackest areas of the vortex. Then, the gray values of the image are inverted, changing black levels into white and the other way around. Then, the gray values of the inverted scale are distributed through the levels contained within the areas, thus magnifying the intensity gradients. The maximum level of both inverted images is shown as a small black area at the center of the Figures. The centroid of these small black corresponds to the center of the vortex. Figure~\ref{fig:VortexSize} shows measurements of the vortex diameter as a function of the driving frequency for different voltages. The vortex grows smaller as the driving frequency rises. The size of the vortex is distinctly the smallest for the lowest voltage $V_1$. Despite having different diameters, the size of the vortices is of the same order of magnitude, varying from about 6 to 16~mm for voltages from $V_2$ up to $V_5$. Vortices stopped forming at frequencies above the last point of each curve of Figure~\ref{fig:VortexSize}. As shown in Figure~\ref{fig:VortexSize}, vortices become smaller with higher frequencies. The displacement, $\Delta x_s$, decreases with the driving frequency as well. This behavior is similar to what has been observed for the stroke ratio \cite{Gharib_JFM1998, Palacios-Morales_AM2013}. However, the stroke ratio in this investigation is significantly larger than in previous studies. According to the LE model in Section~\ref{sec:LEM}, the displacement amplitude is limited by the effects of acoustic inertia, compressibility, and resistance loss.

Measurements of velocity and size of the vortices are combined to estimate the momentum discharged from the nozzle. Such experimental estimation of momentum is compared with results numeric simulations of the LE model. Hereafter, we describe the LE model and its assumptions. 

\section{Nonlinear Lumped Element (LE) model}
\label{sec:LEM}

An acoustic Lumped Element (LE) model, formulated as a nonlinear ordinary differential equation, was used to simulate the resonator dynamics. In this model, the pressure, $P(t)$, and the volume of air in the cavity, $V(t)$, are assumed to vary in time from the initial values of static pressure $P_0$ and volume $V_c$, to the dynamic values
\begin{eqnarray}
	P(t) &=& P_0 + p(t),\\
	V(t) &=& V_c +A_n\,x(t)-V_d(t);
\end{eqnarray}
where $p(t)$ is the sound pressure inside the resonator, $V_d(t)$ is the volume displacement of the inner loudspeaker, $x(t)$ is the oscillatory displacement of air within the neck, and $A_n$ is the inner cross-sectional area of the neck, estimated as $A_n = \pi d_n^2/4$. The inner volume displaced by the driver, $V_d(t)$, is used as a sinusoidal signal input of the LE model
\begin{equation}
	V_d(t) = \sqrt{2} \Delta x_s A_s  \sin (2\pi f t),
\end{equation}
where $A_s$ is the area of the cone of the loudspeaker, $\Delta x_s$ is the RMS amplitude of displacement measured experimentally and shown in Figure~\ref{fig:DispacementVSFrequency}. The area of the loudspeaker cone is $A_s \approx$ 0.0507 m$^2$. The oscillatory displacement in the neck, $x(t)$, is the first unknown to be solved by the simulation. Assuming that the expansion and compression of air in the cavity is  adiabatic~\cite{Singh_JoSV2014,Chiatto_S2017, Rienstra_AIAAJ2018}, changes of pressure and volume satisfy the expression $PV^\gamma $ = $P_0 V_c^\gamma$, where $\gamma$ = 1.4 is the ratio of specific heats at constant pressure and volume. Then, the sound pressure inside the cavity of the resonator can be written as
\begin{equation}
	p(t) = \frac{P_0}{\left(1 + \frac{\Delta V(t)}{V_c}\right)^\gamma} - P_0.
	 \label{eq:SoundPressureResonator}
\end{equation}
The pressure $p(t)$ from Equation~\eqref{eq:SoundPressureResonator} is approximated with the first two terms of the Taylor series as
\begin{equation}
	p(t) \approx -\gamma P_0 \frac{\Delta V(t)}{V_c} + \frac{1}{2}(\gamma+1)\gamma P_0 \left(\frac{\Delta V(t)}{V_c}\right)^2,
	\label{eq:adiabatic-series-expansion}
\end{equation}
where $\Delta V(t)$ = $A_n \,x(t) - V_d(t)$. Since the pressure depends on the displacement, $p(t)$ is the second unknown to be solved by the simulation. To relate the variables of pressure and displacement, we introduced the equation of motion in the neck. The equation of motion for the mass of air $M$ in the nozzle is
\begin{equation}
  M\dot{u} + Ru + \frac{1}{2}\rho_0 u |u| A_n = p(t) A_n,
  \label{Eq:MotionODE}
\end{equation}
where $\rho_0$ is the mean air density, $u = u(t)$ is the instantaneous particle velocity. The particle velocity is estimated as the time derivative of the displacement in the nozzle, $u(t)$ = $\dot{x}(t)$ = $d\,x(t)/dt$. The mass in the nozzle is estimated as $M = \rho_0 A_n L$, where $L$ is the effective length of the nozzle due to the acoustic mass loading, $L = l_n + 0.72 d_n$, assumed approximately as the sum of two standard end corrections: The inner diameter is flanged, 0.41 $d_n$, while the outer diameter is unflanged 0.31 $ d_n$~\cite{Pierce-1981}. The total resistance $R$ in Equation~\eqref{Eq:MotionODE} is the sum of viscous losses on the airflow within the nozzle and sound radiation from the resonator, $R = R_v + R_a$, where both resistances are defined as
\begin{eqnarray}
	R_v = \rho_0 sL(\pi \nu f)^{1/2},
	\label{eq:ViscousLosses}
\end{eqnarray}
and
\begin{eqnarray}
	R_a = \rho_0 c_0A_n (k d_n)^2,
	\label{eq:SoundRadiation}
\end{eqnarray}
where $\nu$ is the kinematic viscosity, $s = \pi d_n$ is the inner perimeter of the nozzle, $k$ = $2\pi f/c_0$ is the acoustic wavenumber.

The quantity $(1/2) \rho_0 u|u|$ in Equation~\eqref{Eq:MotionODE} is the aerodynamic pressure opposing to the pressure in the cavity, behaving as an aerodynamic damper. This term may be derived from the Bernoulli principle and becomes the dominating resistance for sufficiently high amplitudes. Expressions similar to Equation~\eqref{Eq:MotionODE} can be found when modeling Helmholtz resonators \cite{Chiatto_S2017, Rienstra_AIAAJ2018, Singh_JoSV2014} or also in the synthetic-jet literature \cite{Kordik_ET&FSc2017, Gil_JFS2018}. 

\begin{figure}
\centering
\includegraphics[width=0.95\columnwidth]{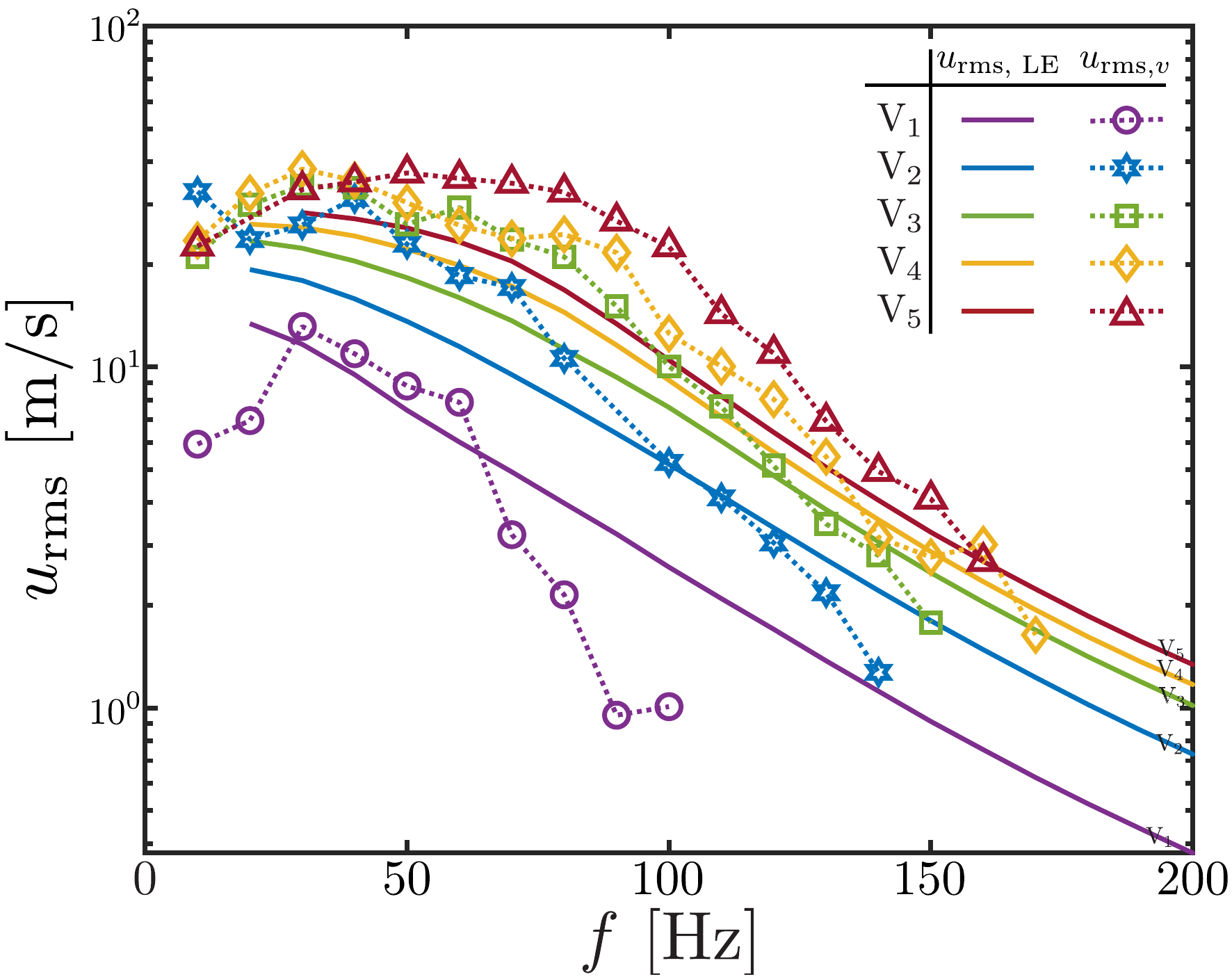}
\caption{\label{fig:ParticleVelocity}{(Color online) RMS amplitude of the particle velocity within the neck. The solid lines correspond to the predictions of the LE model, while the open symbols correspond to the experimental estimation. The particle velocity is solved by feeding the measurements of $\Delta x_s$ to the equation system defined by Equations~\eqref{eq:adiabatic-series-expansion} \& \eqref{Eq:MotionODE}.}}
\end{figure}

Equations~\eqref{eq:adiabatic-series-expansion} and \eqref{Eq:MotionODE} establish a system of equations that are solved simultaneously for the pressure $p(t)$ and the particle displacement in the neck $x(t)$. The solutions were calculated by integrating numerically using a Runge-Kutta routine of combined 4th \& 5th order. This routine is implemented in the function \texttt{ode45} of GNU Octave. Thus, the RMS amplitudes of the pressure $p_\textrm{rms, LE}$ and the particle velocity $u_\textrm{rms, LE}$ were estimated. The values of $p_\textrm{rms, LE}$ are seen as the continuous lines of Figure~\ref{fig:CavityPressure}. For voltages $V_0$ to $V_3$, the experimental results match the results of the LE model. For voltages $V_4$ and $V_5$, the maximum pressure is found at the same frequency as in the experimental results. However, the slope of the pressure is slightly lower than the experimental measurements. The results of the RMS particle velocity $u_\textrm{rms, LE}$ are seen as continuous lines in Figure~\ref{fig:ParticleVelocity}. The results of particle velocity were validated with indirect estimations via the velocity and size of vortices. Such estimations are seen as the open symbols in Figure~\ref{fig:ParticleVelocity}.

\section{Transfer of momentum}
\label{sec:momentum-transfer}

\subsection{Air momentum estimated through the LE model}

The particle velocity calculated from the LE model can be used to estimate the net momentum of air  in a full cycle as
\begin{align}
  \label{eq:MMnet}
  M_\mathrm{net, LE} &= \frac{A_H}{2} \int_{0}^T \rho u|u| \dd{t},
\end{align}
where $A_H$ is the effective hydraulic area of the neck \cite{Kordik_ET&FSc2017}. In this and following sections, the mean density $\rho_0$ has been replaced simply for the density of air $\rho$. Equation~\eqref{eq:MMnet} assumes that air moves from (a) the resonator to the environment when the velocity in the neck is positive, $u>0$, and air moves from (b) the environment to the resonator when the velocity is negative, $u<0$. The momentum estimation in Equation~\eqref{eq:MMnet} could be altered by the geometry of the nozzle. Losses due to flow separation may result in an effective hydraulic area smaller than the inner diameter of the nozzle, $A_H < A_n$. However, the nozzle used in this investigation has an inner chamfer radius of about 1~mm. Also, the length of the nozzle can be considered large, since $l_n / d_n = 5$. Thus, we assumed that the correction due to the discharge coefficient is negligible, resulting in $A_H \approx A_n$. In principle, the expression in Equation\eqref{eq:MMnet} is valid for the steady and incompressible flow in the neck of the resonator. It is assumed that such is a good approximation during time intervals much shorter than the period $\Delta t \sim L_s/u \ll T$, but sufficient for the flow of air through the neck \cite{Pierce-1981}.

\begin{figure}
\centering
\includegraphics[width=0.95\columnwidth]{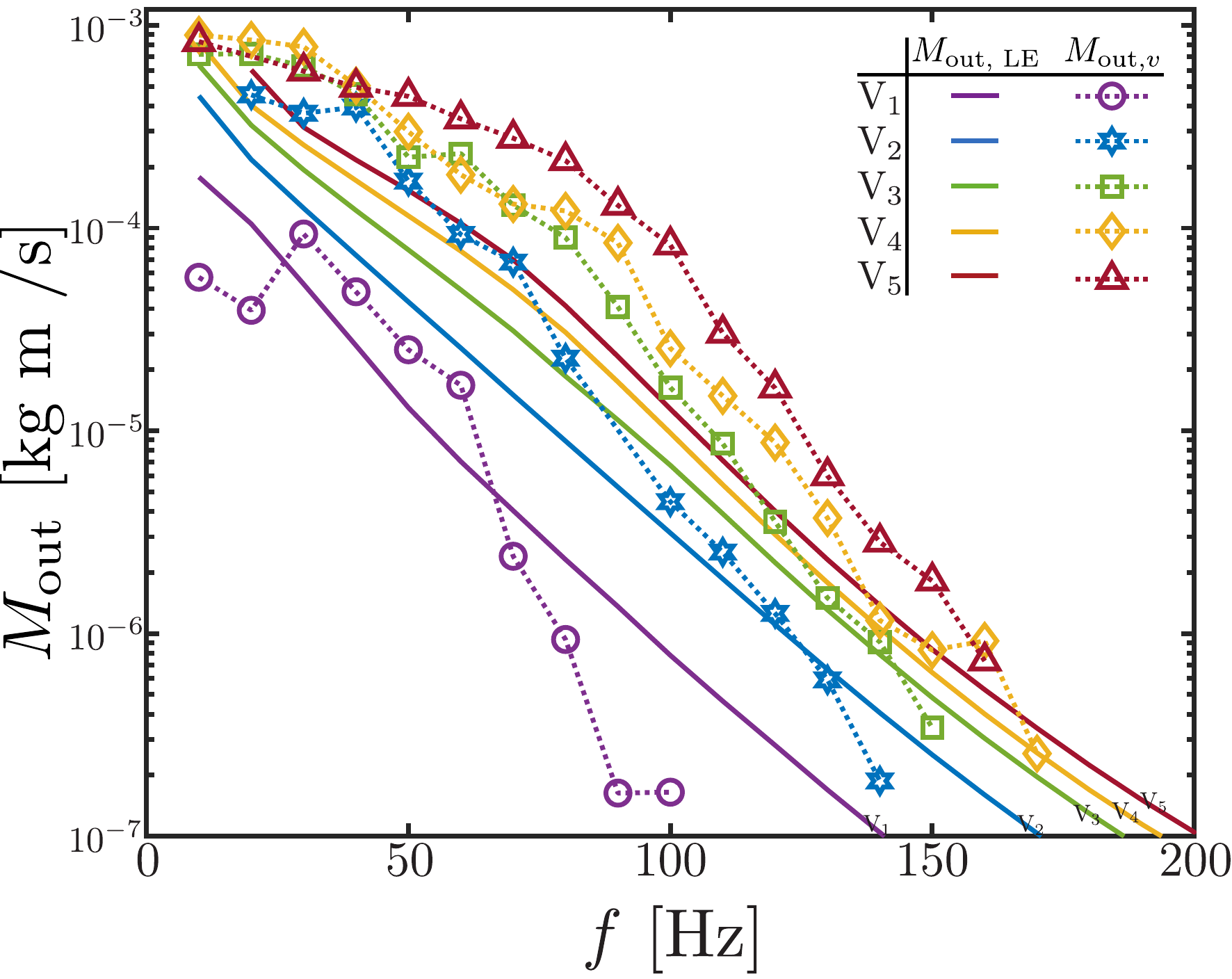}
\caption{\label{fig:MomentumVsF}{(Color online) Momentum transfer per outward cycle as a function of frequency for different voltages. Symbol marks show experimental results, and continuous lines are numerical results of the LE model for $M_\mathrm{out}$.}}
\end{figure}

$M_\mathrm{net, LE}$ has units of linear momentum, mass times velocity, and it is aligned with the axis of the neck. The net momentum transfer can be different from zero because the positive and negative parts of the periodic waveforms of $\rho$ and $u$ are asymmetric.  However, this contribution is not responsible for the main thrust of the external jet. Momentum is transferred to the air only when the particle velocity is positive, rather than during the whole period. Therefore, we define the outflow momentum as the one transferred to the environment asmomentum
\begin{equation}
  \label{eq:M-out}
  M_\mathrm{out, LE} = \frac{A_n}{2} \int_{0, u>0}^T \rho u^2 \dd{t}.
  \end{equation}
For a pure sine wave, the outflow momentum corresponds to the integration in half of the cycle. On the other hand, the inflow momentum is defined as the one transferred back to the device during the negative part of the cycle\begin{equation}
  \label{eq:M-in}
  M_\mathrm{in, LE}  =  -\frac{A_n}{2} \int_{0, u<0}^T \rho u^2 \dd{t}.
\end{equation}
The inflow momentum corresponds to the axial component of the momentum of an inflow converging omnidirectionally from a much wider region of the external fluid. Hence, the momentum transferred per cycle that forms an outer flow and the inflow momentum can be estimated as
where the subscripts $u>0$ and $u<0$ restrict integration to the positive and negative parts of the cycle. 
The underlying assumption is that $M_\mathrm{net, LE} = M_\mathrm{out, LE} + M_\mathrm{in, LE}$. Such cycle decomposition has been previously reported \cite{Gil_JFS2018}. Assuming constant $\rho_0$, the integral in Equation~\eqref{eq:M-out} can be expressed in terms of the RMS particle velocity, as
\begin{equation}
	M_\mathrm{out, LE} = \frac{\pi}{16} \frac{\rho_0 \left({u_\mathrm{rms, LE}}\right)^2 d^2_n}{f}.
	\label{Eq:MomentumLE}
\end{equation}
The solid lines in Figure~\ref{fig:MomentumVsF} show estimations of $M_\mathrm{out, LE}$ as a function of the driving frequency. Equation~\eqref{Eq:MomentumLE} has been implemented previously  \cite{Shuster_PoF2007}.

\subsection{Momentum estimated through visualization}

Measurements of the vortex velocity in Figure~\ref{fig:VortexVel} and their diameter in Figure~\ref{fig:VortexSize} can be additionally used to estimate the momentum discharged through the nozzle. The outflow momentum of the jet can be approximated as $M_{\textrm{out}, v} \approx \rho_0 U_v V_j$, where $V_j$ is the volume of air-jet ejected, estimated as the cross-sectional area of the vortex ring times the stroke length,  $V_j = A_v L_{s,v}$. The cross-sectional area of the vortex ring is estimated with measurements of its diameter, $A_v = \pi D_v^2/4$. Since the jet only develops during the positive cycle, the stroke length is estimated in half a period $L_{s,v} \approx (T/2) U_v = U_v/2f$. Using the visualization measurements, the outflow momentum discharged from the nozzle during the outward cycle is approximated as
\begin{equation}
  M_{\mathrm{out,} v} \approx \frac{\pi}{8} \frac{\rho_0 U_v^2 D_v^2}{f}.
  \label{Eq:MomentumVis}
\end{equation}
These estimations of the outflow momentum $M_{\mathrm{out,} v}$ are shown as open symbols on dotted lines in Figure~\ref{fig:MomentumVsF}. Corrections due to the critical length of vortex formation and other possible effects are not included in the definition of the stroke length $L_{s,v}$. For the lowest intensity, estimations of momentum obtained from visualization measurements are seen as points passing around the prediction of the LE model. However, the momentum is under estimated by the LE model between 51 and 77\% depending on the voltage. Overall, the LE model underestimates the Momentum about a 39\%. We do not find a relationship between the voltage and these estimations.

\subsection{Experimental estimations of particle velocity in the neck}

Momentum estimations from the LE model in Equation~\eqref{Eq:MomentumLE} and the indirect measurements obtained from the visualization  in Equation~\eqref{Eq:MomentumVis} can be compared in Figure~\ref{fig:MomentumVsF}. The voltage labels are kept for the results of the LE model and the experimental estimations because the curves in Figure~\ref{fig:DispacementVSFrequency} are used as input conditions. For the lowest voltage $V_1$, the solid purple line passes through the hollow circles of the experimental estimation. In this case, the LE model over-estimates the velocity about 31\%. However, for voltages from $V_2$ to $V_5$, the LE model underestimated the experimental result between 23 and 33\%, depending on the case. Overall, velocities estimated via visualization were in average 18\% larger than prediction of the LE model. Both estimations followed similar trends and are comparable with one another. Thus, equating Equations~\eqref{Eq:MomentumLE} and~\eqref{Eq:MomentumVis} and solving for the RMS particle velocity leads to an estimation from visualization
\begin{equation}
	u_{\textrm{rms,} v} = \sqrt{2} U_v \left(\frac{ D_v}{d_n}\right).
	\label{Eq:UrmsVis}
\end{equation}
Results from this estimation can be seen as open symbols following a broken line in Figure~\ref{fig:ParticleVelocity}. Similar to Figure~\ref{fig:MomentumVsF}, experimental estimations of the particle velocity and the results of the LE model in Figure~\ref{fig:ParticleVelocity} match fairly for the lowest voltage, and larger deviations are seen for higher voltage curves.

\subsection{Momenta ratio}

\begin{figure}
\centering
\includegraphics[width=0.95\columnwidth]{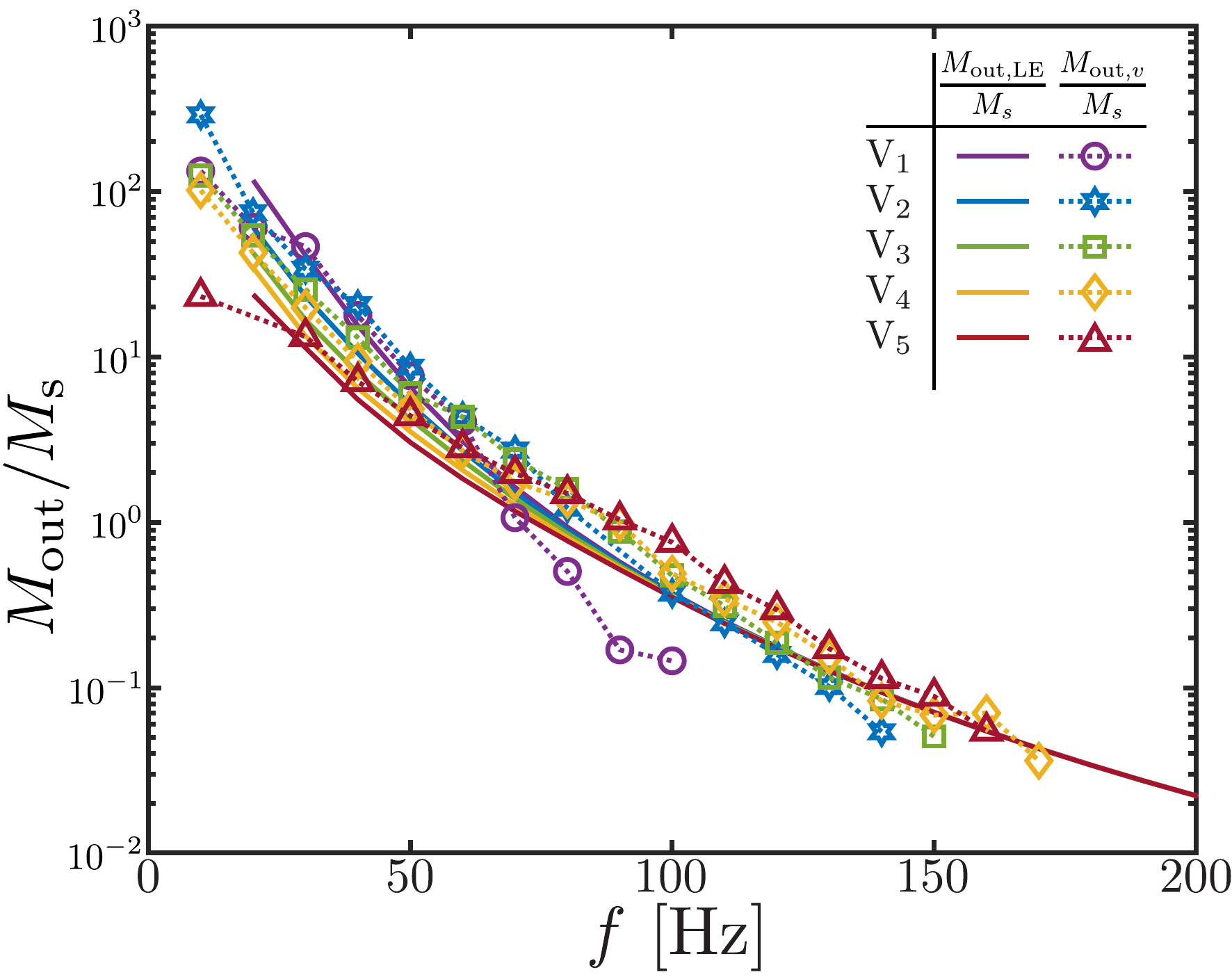}
\caption{\label{fig:MomentumRatio}{(Color online) Outflow momentum $M_\textrm{out}$ normalized by the input momentum of the speaker $M_s$. The ratio $M_\textrm{out}/M_s$ is plotted as a function of the frequency comparing two estimations, the one from the LE model and the visualization.}}
\end{figure}

To give a dimensionless estimation of the outflow momentum, we propose to normalize it using a reference value of the input momentum transfer of the loudspeaker, defined in terms of the RMS velocity and diameter of the speaker
\begin{equation}
	M_\mathrm{s} = \frac{\pi}{16} \frac{\rho_0 {v_s}^2 d^2_s}{f},
	\label{Eq:MomentumSpeaker}
\end{equation}
where $v_s$ was estimated experimentally as defined in Equation~\eqref{Eq:RMSVelocity}. Estimated from the LE model, the momenta ratio of outflow to input is
\begin{equation}
	\frac{M_\textrm{out, LE}}{M_s} = \left(\frac{u_\textrm{rms, LE}\; d_n}{v_s d_s}\right)^2.
	\label{Eq:MomentumRatioLEmodel}
\end{equation}
The solid lines in Figure~\ref{fig:MomentumRatio} show the estimation of momenta ratio of Equation~\eqref{Eq:MomentumRatioLEmodel}. The same approach can be used to estimate the momenta ratio from the experimental visualization, leading to
\begin{equation}
	\frac{M_{\textrm{out}, v}}{M_s} =  \left(\frac{u_{\textrm{rms,} v}\; d_n}{v_s d_s}\right)^2.
	\label{Eq:MomentumRatioVisualization}
\end{equation}
The open symbols of Figure~\ref{fig:MomentumRatio} show the experimental estimation of Equation~\eqref{Eq:MomentumRatioVisualization}. As seen in Figure~\ref{fig:MomentumRatio}, the experimental estimation and the results from the LE model follow the same trend and match fairly with one another. Similar to Figure~\ref{fig:MomentumVsF}, the LE model underestimates the Momenum ratio in some cases of Figure~\ref{fig:MomentumRatio}.

The momenta ratio $M_\textrm{out}/M_s$ in Figure~\eqref{fig:MomentumRatio} reaches values above one for frequencies lower than $f$ = 50 Hz. This result is predicted by the LE model and confirmed by the experimental measurements. A harmonic oscillator driven by a forced vibration may reach transmissibility larger than one in some range of frequencies around resonance. The reason for this is that reactive forces, such as inertia and elasticity, cancel each other, leading to a significant increase in the amplitude in the neck. Such an increase may reach output amplitudes larger than the input.

Similar to the estimations of momentum, the momenta ratio is under-predicted by the LE model by a factor smaller than two in all the cases. However, the change does not seem so significant because the momenta ratio varies more than 3 decades.
\subsection{Stroke length}
\begin{figure}
\centering
\includegraphics[width=0.95\columnwidth]{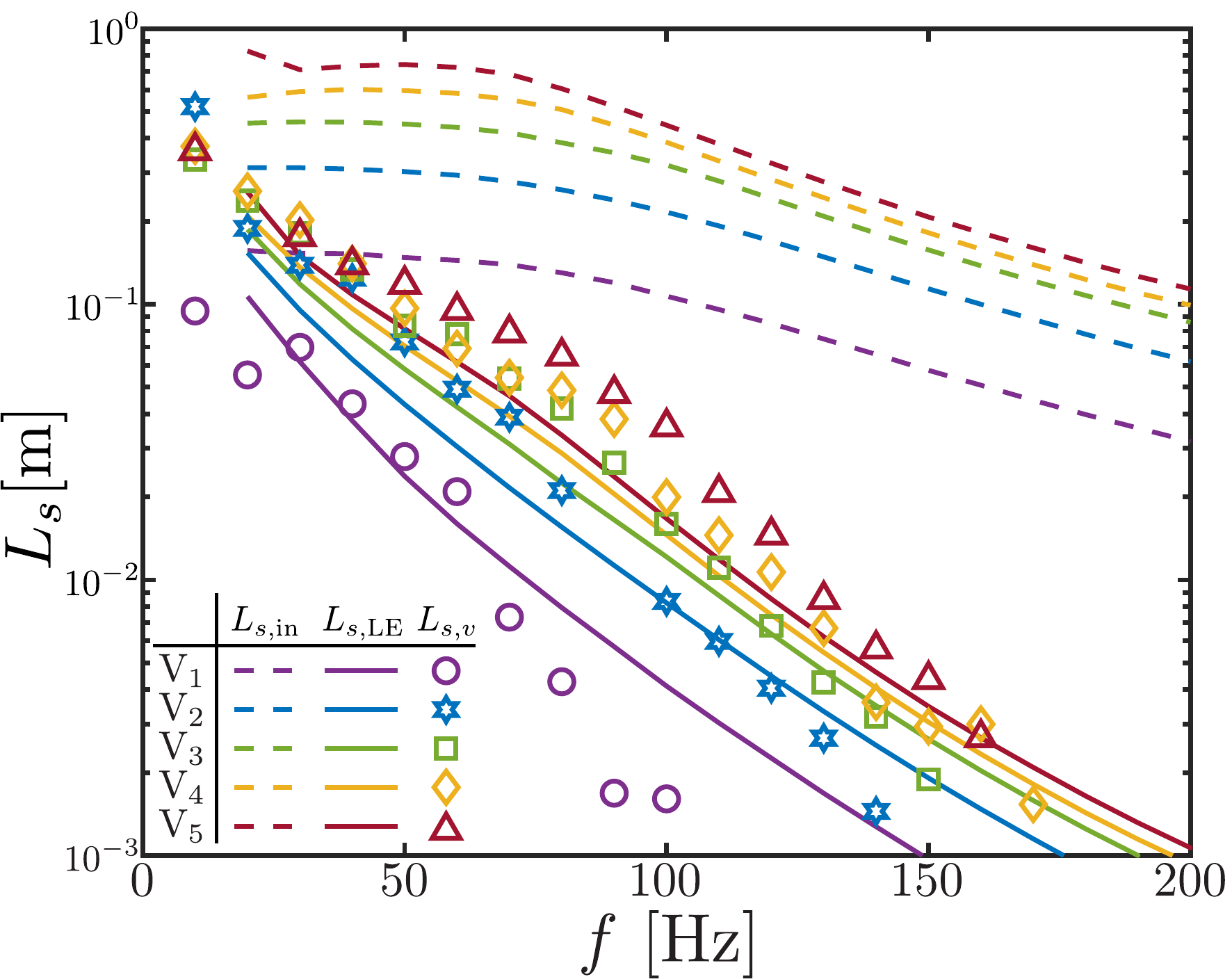}
\caption{\label{fig:StrokeLength}{(Color online). Stroke length as a function of the driving frequency estimated with three different assumptions. (a) Estimations of the broken lines assume incompressibility. (b) Estimations obtained from the LE model are seen as solid lines. (c) Estimations from the experimental visualization are shown as open symbols.}}
\end{figure}

At first glance, one could think that the flow could be modeled as incompressible. If this were the case, the incompressible stroke length could be estimated as the RMS displacement of the loudspeaker corrected by the ratio between the area of the speaker and the neck of the resonator
\begin{equation}
	L_{s, \textrm{in}} = \Delta x_s \frac{A_s}{A_n} = \Delta x_s \left(\frac{d_s}{d_n}\right)^2,
	\label{eq:StrokeLengthIncompressible}
\end{equation}
where $\Delta x_s$ was determined experimentally as defined in Equation~\eqref{Eq:RMSDisplacement}. Curves estimated with Equation~\eqref{eq:StrokeLengthIncompressible} are shown as solid lines in Figure~\ref{fig:StrokeLength}. However, the values of sound pressure, estimated in Figure~\ref{fig:CavityPressure}, are large enough to produce considerable compressibility within the cavity of the resonator. Thus, the stroke length may also be estimated using the LE model and the visualization experiments. Using the definition of Equation~\eqref{eq:StrokeLength}, the stroke length estimated from the LE model is
\begin{equation}
	L_{s, \textrm{LE}} = \frac{u_\textrm{rms, LE}}{2\pi f}.
	\label{eq:StrokeLengthLE}
\end{equation}
The velocity scale used in Equation~\eqref{eq:StrokeLengthLE} is the RMS particle velocity estimated from the LE model, $u_0 = u_\textrm{rms, LE}$. Curves estimated using Equation~\eqref{eq:StrokeLengthLE} are shown as broken lines in Figure~\ref{fig:StrokeLength}. Similarly, the stroke length can be estimated with the visualization experiments as
\begin{equation}
	L_{s,v} = \frac{u_{\textrm{rms,} v}}{2\pi f}.
	\label{eq:StrokeLengthVis}
\end{equation}
Estimations of Equation~\eqref{eq:StrokeLengthVis} are shown as the open symbols in Figure~\ref{fig:StrokeLength}. Assuming the flow is incompressible results in a considerable overestimation of the stroke length. The effects of compressibility play a significant role in the estimation of the stroke length because the values of sound pressure in the cavity, reported in Figure~\ref{fig:CavityPressure}, are significantly high. Thus the dynamics within the cavity should be considered highly compressible, unlike in other synthetic jet devices.

Since the stroke length scales linearly with the velocity, the under-estimation of the LE model conforms to the same values as the particle velocity. This results in a maximum under-estimation of 33\% but its mean is reduced to 18\%.

\subsection{Reynolds and Strouhal numbers}
The dynamics of synthetic jet devices and other resonators have been typically characterized with the Reynolds and the Strouhal numbers. By studying simultaneously the flow within the neck of the resonator and the outflow, we have connected two very different regimes. On the one hand, the regime of flow can be studied with the Reynolds number within the neck of the resonator as
\begin{equation}
	\textrm{Re}_n = \frac{\rho_0 \, u_\textrm{rms, LE} \, d_n}{\mu},
	\label{Eq:ReynoldsNeck}
\end{equation}
where the velocity scale has been defined as $u_0 = u_\textrm{rms, LE}$. On the other hand, the Reynolds number can be defined for the outflow using the velocity and size of the vortex
\begin{equation}
	\textrm{Re}_v = \frac{\rho_0 \, U_v D_v}{\mu}.
	\label{Eq:ReynoldsVortex}
\end{equation}
From Equation~\eqref{Eq:UrmsVis} it can be seen that the product $u_\textrm{rms, LE} \, d_n$ in Equation~\eqref{Eq:ReynoldsNeck} is related to $U_v D_v$ in Equation~\eqref{Eq:ReynoldsVortex} by a factor $\sqrt{2}$. However, even if the definitions are similar, the Reynolds numbers refer to different phenomena. Two modified Strouhal numbers can be defined in a similar way. In the neck, the Strouhal number can be defined as a function of the RMS velocity
\begin{equation}
	\textrm{Sr}_n = \frac{f d_n}{u_\textrm{rms, LE}}.
	\label{Eq:StrouhalNeck}
\end{equation}
In the outflow, another Strouhal number can be defined as a function of the velocity and size of the vortex
\begin{equation}
	\textrm{Sr}_v = \frac{f D_v}{U_v}.
	\label{Eq:StrouhalVortex}
\end{equation}
Unlike the definitions of the Reynolds numbers, the ratio $d_n/u_\textrm{rms, LE}$ in Equation~\eqref{Eq:StrouhalNeck} is not related by a constant factor to $D_v/U_v$ in Equation~\eqref{Eq:StrouhalVortex}. Here, it should also be taken into account that $d_n$ is a constant, while the size of the vortex $D_v$ decreases with higher frequencies.
\begin{figure}
\centering
\includegraphics[width=0.95\columnwidth]{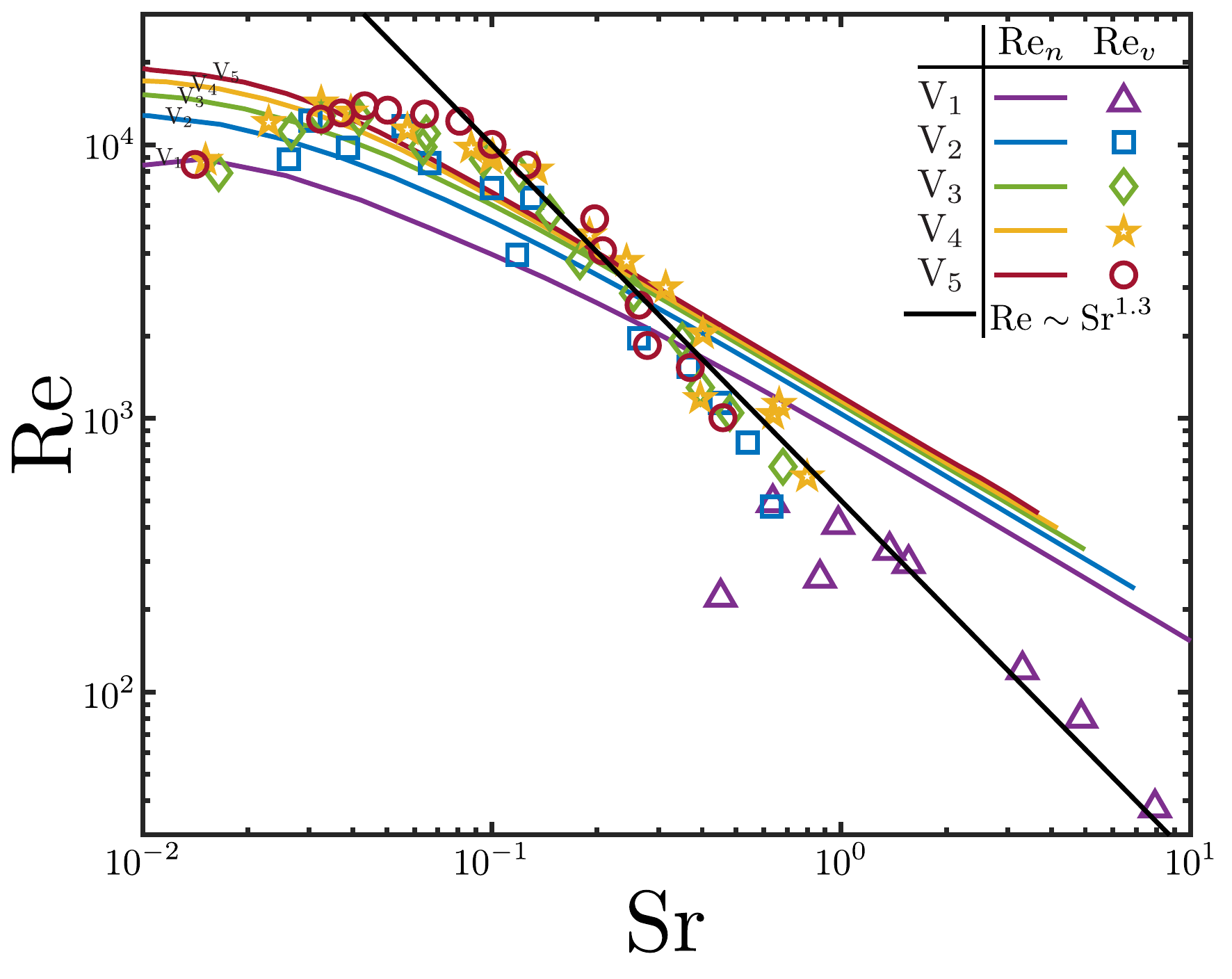}
\caption{\label{fig:ReVSSr}{(Color online). Reynolds number plotted as a function of the Strouhal number. The solid lines correspond to the results of the LE model within the neck of the resonator. The open symbols correspond to results of the experimental visualization of the outflow.}}
\end{figure}
The solid lines in Figure~\ref{fig:ReVSSr} show the estimations of the Reynolds and the Strouhal numbers within the neck. The open symbols in Figure~\ref{fig:ReVSSr} show the same estimations for the outflow. The solid black line is an empirical relationship scaling as Re $\sim$ Sr$^{1.3}$. Even if the values of Reynolds are similar for the lowest Strouhal numbers, the solid lines and the open symbols in Figure~\ref{fig:ReVSSr} display different decay exponents. Furthermore, for low Strouhal numbers, the solid lines seem to separate in different curves. But, for the outflow, the symbols seem to have collapsed on the entire range. The different slopes describe a different behavior of velocity vs frequency in the outflow and the neck of the resonator.

The decay curves shown in Figure~\ref{fig:ReVSSr} obtained by the LE model do not collapse to a master curve. This behavior is consistent with previous findings where Synthetic jets operate in a wide range of independent conditions. However, the experimental points seem to overlap in the decay region. Since the geometry of the SJ and the properties of the active fluid remain the same, we are not in the position to state that this dynamic collapse is universal. The experimental points depend on two parameters selected by the system dynamics: the vortex velocity and its diameter, unlike the momentum transferred from the neck.


\section{Discussion}

The flow of the synthetic jet device should be modeled using the compressible Navier-Stokes equations at the high SPL reported in Table ~\ref{Tab:VoltageValues}. However, the LE model condenses the physics of the flow into an ordinary differential equation. Thus, an estimation of the mean particle velocity within an 18\% of error is reasonable. Even the highest deviation, 33\%, remains within the same order of magnitude. For the highest amplitudes, the LE model underpredicts the particle velocity in the neck. This result underlines that the total resistance $R$ in Eq.~\ref{Eq:MotionODE} overestimates losses due to viscous or/and sound radiation effects. Since the underprediction ceases at the lowest SPL, we may argue that $R$ is dynamic. However, identifying the mechanisms responsible for such a behavior are beyond the scope of this manuscript. Changes in the neck geometry are to be studied further in future work.

The apparent collapse of the experimental data in Figure~\ref{fig:ReVSSr} is by no means universal. Different geometries of the cavity, the neck, and properties of the active fluid, \textit{e.g.}, viscosity,  density should be tested.

Experimental measurements of sound pressure in the resonator's cavity and corresponding predictions of the LE model, seen in Figure~\ref{fig:CavityPressure}, are surprisingly similar. The behavior at low frequencies and up to the maximum pressure tallies for every voltage. The trend of all the curves on high frequencies converges to the same values as well. However, for voltages $V_4$ and $V_5$ on higher frequencies than the non-linear resonance, the LE model predicts a higher slope decline than the experimental measurements. A different decline rate of the pressure may yield different slopes in the particle velocity decay in Figure~\ref{fig:ParticleVelocity} or the stroke length in Figure~\ref{fig:ReVSSr}. However, the most evident difference between the prediction and the experiments manifests when comparing the Reynolds and the Strouhal numbers in Figure~\ref{fig:ReVSSr}. Such a distinct behavior cannot be attributed only to differences between the model and the experiments, but Reynolds and the Strouhal numbers describe distinctive phenomena. While the dimensionless numbers on the neck outline the regime of the inner flow, $Re_v$ and $Sr_v$ characterize the outflow dynamics. As vortices and jets form in the outflow, some energy conversion is occurring halfway. Such a transformation may underlie an energy loss that we do not have the means to quantify at the time.

We believe that underestimations of Momentum and particle velocity appear because the LE model under estimates the transmissibility of the system. However, our results are not sufficient to identify the mechanism responsible. Despite such a discrepancy, Figure~\eqref{fig:MomentumRatio} shows that the momentum transfer ratio $M_\textrm{out}/M_\textrm{s}$ differs almost by a constant value and behaves consistently among the measured and simulated results obtained for the different driving amplitudes.

The conditions that promote the mechanism of momentum transfer under study are: (1)~An outward jet is formed in the outflow, which requires a large Reynolds number, $\Re_n \gg 1$ and $\Re_v \gg 1$. Since the values of the Reynolds number were comprehended between $3.6\times10^{2}$ and $1.4\times10^{4}$, this condition is fulfilled. Also, this is a condition for negligible viscous damping in the neck of the resonator, and large aerodynamic resistance. (2)~A sufficiently large stroke is manifested in the outflow half-cycle, relative to the dimensions of the neck, requiring a small Strouhal number, $\Sr_n  \ll 1$ and $\Sr_v  \ll 1$. In both the neck and the vortex, the Strouhal number varied from $4\times10^{-2}$ to $\times10^{1}$. These can be combined in terms of the ratio between the Reynolds and the Strouhal numbers into the condition: $\Re / \Sr = (d_n / l_n)\ (u_\textrm{rms}^2 / \nu f) \gg 1$.

The role of the Strouhal number lies in the dimensionless stroke, while the Reynolds quantifies the ratio of dynamic pressure to the stress on the inner wall of the neck. The stress on the wall depends on the viscosity and the length of the neck of the resonator. None of these two parameters was varied in this study. In the investigation of Tesa\u{r} \textit{et al.} (2006), the stress on the wall was varied by having longer loop lengths. These lengths were varied from 1 up to 52~m. Investigating such a large variations, resulted in curves with different slopes relating the Reynolds and the Strouhal numbers. Thus, these two numbers were demonstrated to be independent of one another and both dominate the dynamics of the flow \cite{Tesar_S&AA2006}. Unfortunately, the momentum transferred per cycle is not a quantity comprising all the aspects of the resonator dynamics and therefore, these results are not scaled using these dimensional numbers.

Borda-Carnot corrections may limit the estimations of momentum transferred in Equation~\eqref{eq:MMnet}. These corrections result from a reduction in the effective hydraulic area due to the formation of re-circulation zones close to the sharp edge of the contraction. The prefactor is known as the discharge coefficient, $c_D = A_H/A_n$. the discharge coefficient may drop down to $c_D$ = 0.5 for very short lengths of the nozzle, $l_n/d_n\ll 1$. But, $c_D \approx$ 1 when the nozzle has round edges and can be considered long ($l_n/d_n \geq$ 1). In this investigation, the nozzle can be considered long, $l_n/d_n = 5$, and its edges are rounded with a chamfer radius of about 1~mm. In previous characterizations, these two conditions usually result in $c_D > 0.9$. The Bernoulli velocity can be estimated as a function of the pressure, $u_\textrm{rms} = f_{BC} \sqrt{p_\textrm{rms}/\rho_0}$, where $f_{BC}$ is the Borda-Carnot factor that corrects the velocity, written as
\begin{equation}
f_{BC} = \sqrt{\frac{2}{1+\frac{1}{2 c_D^2} -\frac{1}{c_D}}}.
\end{equation}
In the absence of losses, the discharge coefficient is $c_D = 1$ and the Borda-Carnot factor is the traditional result in the Bernoulli equation, $f_{BC}$ = 2. For the maximum losses, we assumed that the discharge coefficient is $c_D$ = 0.9, resulting in a Borda-Carnot factor of $f_{BC}$ = 1.9878. Then, using $A_n$ in Equations~\eqref{eq:MMnet}, \eqref{eq:M-out} and \eqref{eq:M-in} to estimate the momentum transferred per cycle leads to an overestimation of about 1\%. Such a correction is smaller than the changes we observe. This explains why the numeric simulations and the experimental estimations of momentum and the dimensionless momentum transferred per cycle match fairly with one another. Since the discharge coefficient varies with the geometry of the nozzle, the surface roughness and the Reynolds number, providing specific values would require an independent study.

The outward flow can be quantitatively estimated by evaluating the outer momentum $|M_\mathrm{out}|$ in the neck of the resonator using Equation~\eqref{eq:M-out}.
The vortex formation during the outward cycle results in linear momentum condensed in a relatively small area immediately outside the nozzle. During the inward cycle, most of the momentum is balanced back $|M_\mathrm{in}| \approx |M_\mathrm{out}|$. But, the momentum returns from a much wider external region upstream the nozzle because the contraction changed its orientation and no particular flow structures are formed outward the nozzle. As vortices form due to viscous effects on the wall of the nozzle, viscosity is responsible for the asymmetric distribution of momentum inside and outside of the nozzle at any given moment.


\section{Conclusions}

We used a LE model based on a nonlinear differential equation, to predict the momentum transferred from a Helmholtz resonator to the environment. Measurements of the RMS displacement, $\Delta x_s$, are the signal input of the LE model. Predictions of pressure within the cavity of the resonator match fairly with measurements of the RMS values. While predictions of particle velocity in the neck are used to estimate the momentum transferred per cycle. Momentum discharged from the neck is modeled assuming that an external jet is produced when the instantaneous velocity is positive, $u > 0$, and larger than a threshold. Such predictions match fairly with the estimations of momentum carried out with the velocity and size of the vortices formed in the external flow. Measuring the velocity and size of vortices, we estimated the particle velocity in the neck of the resonator. Such an optical estimation of the particle velocity shows comparable results with the LE model. Despite the differences between the estimations of particle velocity in Figure~\ref{fig:ParticleVelocity}, the estimations of momentum ratio in Figure~\ref{fig:MomentumRatio} are very similar.

A stationary Helmholtz resonator driven at high amplitudes drives out a mean external outflow. Such flow is observed as jets and vortices of different size and velocity.
The outward transfer of momentum in the outflow cycle, estimated from flow visualization, is consistent with results from numerical simulations of the outflow cycle in the neck of the resonator. 
Conditions leading to the transfer of momentum in the outflow cycle require (1)~a large Reynolds number $\Re = u_\textrm{rms} d_n / \nu \gg 1$, promoting the formation of an external jet, a small effect of viscous damping in the neck of the resonator, and a large aerodynamic resistance; and (2)~a small Strouhal number $\Sr = f l_n / u_\textrm{rms}$, for a sufficiently large outflow stroke, relative to the dimensions of the neck.
A non-dimensional scale favorable for momentum transfer in the outflow cycle in this type of device can be established as the ratio of the Reynolds and the Strouhal numbers: $\Re / \Sr = (d_n / l_n)\ (u_\textrm{rms}^2 / \nu f) \gg 1$.

\section*{Appendix}
\subsection*{Vortex width measurement method}

\begin{figure}
\centering
\includegraphics[width=0.95\columnwidth]{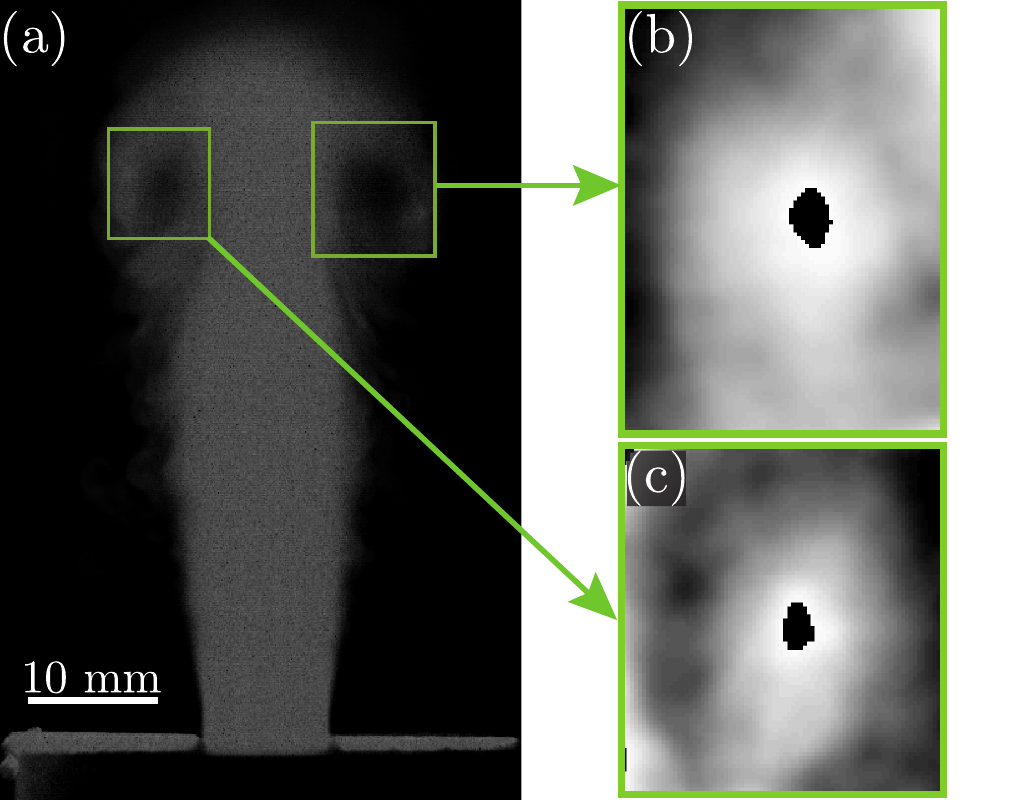}
\caption{\label{fig:VortexWidthMeasurement}{(Color online). (a) Average image of a vortex ring. The vortices were driven at 20 Hz and . The experiment was recorded at 2600 fps. The figure shows the entire external diameter of the nozzle as a horizontal bright line. (b) A close up to the right side of the average image. (c) A close up to the left side of the average image. Both close ups show an inverted image with the gray values distributed within contained in the green area. The small black areas at the center of (a \& b) show the maximum of the inverted image.
}}
\end{figure}

The diameter of the vortex was estimated as the distance between the two centers seen in the average image of each experiment. An example of an average image is presented in Figure \ref{fig:VortexWidthMeasurement}(a). The average image was estimated using about 10 independent frames at the same point of the cycle. Since the center of the vortex is not clearly seen in the average image, we have developed the following procedure to estimate more precisely the centroid of the vortex. The green rectangles in Figure \ref{fig:VortexWidthMeasurement}(a) are overseen as inverted gray scale images in Figures \ref{fig:VortexWidthMeasurement}(b \& c). The gray values of the inverted scale are distributed through the levels contained within the green areas, thus magnifying the gradients. The maximum level of both inverted images is shown as a small black area at the center of Figures \ref{fig:VortexWidthMeasurement}(b \& c). The centroid of these small black corresponds to the center of the vortex.

\subsection*{Vortex velocity measurement method}

The Videos S1 \& S2 show the formation and evolution of vortices in two different conditions. The driving frequency in Video S1 was 30 Hz and for Video S2 it was 150 Hz. Video S1 was recorded at 2600 fps and Video S2 was recorded at 2000fps. Both videos are being reproduced at 5 fps. The external diameter of the nozzle is visible as a discontinuous horizontal line. Its outer diameter measures 40 mm. The horizontal yellow line in both videos exemplify the result of manually tracking the position of the vortices. The position of the vortex front is measured in at least 7 frames and recorded. Then, the position and time are fitted to a straight line and the velocity  is estimated as the slope of such fit. The values of  are estimated as the average of at least 10 vortices and 3 different experiments.

\section*{Acknowledgments}
This research was supported by DGAPA UNAM PAPIIT, grant numbers IA102220, and TA100620.
Technical support for the experimental work was provided by Antonio P\'erez-L\'opez and Ricardo Dorantes-Escamilla. The authors would like to thank Pablo Rend\'on, Roberto Zenit, and Salvador S\'anchez Minero for their support and the insightful discussions on this project.

\bibliographystyle{elsarticle-num-names}
\bibliography{Manuscript}

\end{document}